\documentclass[12pt]{article}
\usepackage{amssymb,amsmath,amsthm,fullpage,graphicx,bm,float}
\newtheoremstyle{custom}{3pt}{3pt}{}{}{\bfseries}{:}{.5em}{}

\title{Optimal mixing enhancement}
\author{Gary Froyland\\ School of Mathematics and Statistics\\ University of New South Wales\\ Sydney NSW 2052, Australia\\ \\ Naratip Santitissadeekorn \\ Department of Mathematics \\ University of Surrey \\ Guildford, GU2 7XH
United Kingdom}
\date{26 March 2017}
\begin{document}
\maketitle
\begin{abstract}
We introduce a general-purpose method for optimising the mixing rate of advective fluid flows.
An existing velocity field is perturbed in a $C^1$ neighborhood to maximize the mixing rate for flows generated by velocity fields in this neighborhood.
Our numerical approach is based on the infinitesimal generator of the flow and is solved by standard linear programming methods.
The perturbed flow may be easily constrained to preserve the same steady state distribution as the original flow, and various natural geometric constraints can also be simply applied.
The same technique can also be used to optimize the mixing rate of advection-diffusion flow models by manipulating the drift term in a small neighborhood.
\end{abstract}
\section{Introduction}

Mixing in fluids is a question of fundamental interest in engineering, the physical sciences, and the natural sciences, with applications ranging from industrial and chemical mixing on small and large scales, to preventing the spreading of pollutants in geophysical flows.
In many applications (e.g.\ microfluid mixing~\cite{Balasuriya15})
homogeneity requirements for mixtures are becoming more stringent as mixing technology improves and new materials are developed.
The goal of this work is to introduce a general purpose method for manipulating the mixing of fluids, under user-specified limits on the extent of the external changes to the generating dynamical flow.
We refer the reader to the book of Sturman \emph{et al.} \cite{SturmanEtAl} for mathematical foundations of advective (kinematic) mixing, and to \cite{ChaoticAdvectionReview} for a recent survey of chaotic advection.

For $M\subset \mathbb{R}^d$ a compact, smooth manifold with vanishing\footnote{the obvious extensions can be made for general Riemannian manifolds.} curvature and $v:M\to\mathbb{R}^d$ a steady\footnote{Periodically driven flows will be discussed in Section \ref{sect:periodic}.} vector field, the Fokker-Planck equation (or Kolmogorov forward equation or advection-diffusion equation)
\begin{equation}
\label{FPE}
\frac{\partial u}{\partial t}=-\nabla\cdot(vu)+\frac{\epsilon^2}{2}\triangle u,
\end{equation}
describes the evolution of a concentration field $u(x,t)$ in the fluid.
The right hand side of (\ref{FPE}) is a second order elliptic operator $\mathcal{A}$ (the infinitesimal operator or infinitesimal generator\footnote{We remark that the word ``generator'' is also used for the infinitesimal operator associated with the Kolmogorov backward equation.  This operator is the $L^1-L^\infty$ adjoint of our generator $\mathcal{A}$;  see e.g.\ \cite{LasotaMackey} for details.}).
We assume conservation of mass, namely zero Neumann boundary conditions, $v(x)\cdot \nabla n(x)=0$ for all $x\in\partial M$, where $n(x)$ is the outward pointing normal vector at $x\in\partial M$;  this ensures the spectrum of $\mathcal{A}$ lies in the left half of the complex plane and includes $0\in \mathbb{C}$.
For steady flows, the eigenfunction corresponding to the zero eigenvalue describes the steady state concentration field.
The rate at which initial concentrations converge to this equilibrium concentration depends on the gap between the imaginary axis and the real part of the next spectral value.

An important global feature of mixing, \textit{strange eigenmodes} \cite{Pierrehumbert94}, arises directly from eigenfunctions of the infinitesimal generator.
Strange eigenmodes are eigenfunctions of the infinitesimal generator that correspond to eigenvalues close to the imaginary axis, when the remainder of the spectrum is much farther from the imaginary axis;  that is, there is a large spectral gap between the two collections of eigenvalues.
This large spectral gap means that the strange eigenmodes decay much more slowly and survive over much longer timescales than lower eigenmodes;  see \cite{FJK13} for a discussion in the present setting and \cite{DellnitzJunge99,DFS00,F07} for analogous results in discrete time. The term ``strange" was suggested since the spatial scale of the eigenmode can be arbitrarily small as the diffusion approaches zero.
When the flow is time-periodic, the strange eigenmodes are themselves time-periodic \cite{Voth,FrKo};  see also \cite{PikovskyPopovych} and \cite{FroylandLloydSanti} for the discrete time case with periodic and aperiodic driving, respectively. At low P\'eclet numbers ($\epsilon>0$ small), level sets of the strange eigenmodes correspond to parts of the domain $M$ that are almost-invariant \cite{FJK13} (resp.\ almost-periodic \cite{FrKo}) under the steady flow $\dot{x}=v(x)$ (resp.\ under the periodically driven flow $\dot{x}=v(x,t)$).
If the goal is to enhance mixing, then one of the outcomes of our approach will be to weaken the invariance properties of these structures.
The concept of strange eigenmodes has been applied to the identification of oceanic gyres \cite{FPET,FSvS14}, and the tracking of atmospheric vortices \cite{FroylandSantiMonahan} and ocean eddies \cite{FroylandAgulhas}. Strange eigenmodes have also been investigated in the context of open flows \cite{GouillartDauchotThiffeault,GouillartDauchotThiffeaultRoux} and flows in Batchelor-regime turbulence, where the momentum diffusivity is much larger than the scalar diffusivity~\cite{FH04,HV05}.

In prototype flows of industrial mixing and static mixing devices, a~\textit{mapping method} introduced in \cite{SpencerWiley} has been recently applied to analyse, design and optimize mixing protocols~\cite{SinghSpeetjensAnderson,SinghKangMeijerAnderson,GorodetskyiSpeetjensAnderson} for time-periodic flows.
We remark that the mapping method \cite{SpencerWiley} is equivalent to Ulam's method \cite{Ulam}, which we introduce later.

Before one can optimize mixing, one first needs to quantify mixing.
Popular mixing measures in the advection-diffusion setting have been based on quantities such as dispersion statistics and spatial variance of passive scalar concentrations; see e.g. \cite{Provenzale,LiuHaller,ThiffeaultDoeringGibbon,ThiffeaultChapter}.
Mathew, Mezi\'c and Petzold \cite{MathewMezicPetzold} introduced a {multiscale measure for mixing} on tori.
This multiscale norm, called the mix-norm, is equivalent to a Sobolev norm of index $-1/2$.
If one instead uses an index of $-1$, the decay of a concentration field in the corresponding Sobolev norm is equivalent to weak convergence of the concentration field to zero in $L^2$, provided the concentration field remains uniformly bounded in $L^2$ for all $t>0$ \cite{Thiffeault}.
Further, if one obtains the above weak convergence for all $f\in L^2$ with zero mean, then this is equivalent to the notion of mixing in ergodic theory \cite{Walters,Petersen}.
We refer the reader to \cite{Thiffeault} for a detailed review on multiscale norms, including an extensive list of references.

Our approach in the present work is to quantify the (exponential) mixing rate as the largest nontrivial real part of the spectrum of the generator;  that is,
\begin{equation}
\label{Lambdaeqn}
\Lambda:=\max\{\Re \lambda: \lambda\in \sigma(\mathcal{A})\setminus\{0\}\}<0.
\end{equation}
If $u(\cdot,0)$ is a concentration field with  zero mean, we have
\begin{equation}
\label{L2mix}
\|u(\cdot,t)\|_{L^2}\le Ce^{\Lambda t}\mbox{ for all $t>0$},
\end{equation}
where $C$ is a constant depending on $u(\cdot,0)$, but the \emph{rate $\Lambda$ is independent of $u(\cdot,0)$}.
Thus, one has that $u(\cdot,t)$ converges \emph{exponentially quickly (at rate $\Lambda$) to zero in $L^2$ norm}.

With these established methods of quantifying mixing, one can now turn to the problem of \textit{optimising} mixing.
Existing methods include switching the vector field between a small number of prespecified vector fields \cite{Mathew_Mezic_Grivopoulos_Vaidya_Petzold,Cortelezzi_Adrover_Giona08, Padberg-Gehle_Ober-Blobaum}, or by additionally optimising a small number of parameters for the vector field \cite{Gubanov_Cortezzi10}.
In contrast to the above work we carry out a \emph{full} optimisation over all vector fields in a small neighborhood of the provided vector field, rather than a restricted low-dimensional optimisation based on switching between specified vector fields or tuning a small number of control parameters.
The problem we solve is more difficult because of its high dimensionality, but the payoff is that we can find more efficiently mixing vector fields because we do not restrict ourselves to a handful of prespecified vector fields.
Another crucial difference between our current approach and all previous approaches mentioned above is that our proposed optimisation provides a guaranteed mixing rate for \emph{all} initial concentration, and is not only effective for a specific initial concentration.
This advantage arises from the manipulation of the \emph{spectrum} of the generator, rather than the manipulation of \emph{norms of specific concentration fields}.

There is analytic work on constructing rapidly mixing velocity fields, including \cite{LinThiffeaultDoering,ACM,YZ};  these methods do not consider perturbations or tunings of a given velocity field.
Further analysis of the rate of mixing achieved in \cite{LinThiffeaultDoering} is provided in \cite{LLetal} and \cite{IKX} also provide bounds on achievable mixing rates for enstropy-limited velocity fields.
As an alternative to optimising the velocity field, \cite{FGTW} determine an optimal local diffusion to minimize the $L^2$ distance between the current concentration field and the steady state in one time step.
Other work on fluid mixing includes \cite{ThiffeaultPavliotis} who identify optimal source distributions which are best mixed for given stirring field and diffusivity.
Balasuriya \cite{Balasuriya15} surveys dynamical systems techniques to enhance mixing in the context of microfluids.





Our main contribution in this paper is to numerically determine an optimal small perturbation of a given vector field which most enhances mixing.
This goal can be easily reversed to find an optimal small perturbation that \emph{most inhibits mixing}, and further, one can also \emph{target specific strange eigenmodes} to either weaken or strengthen their effect on overall mixing.
Our approach is based on representing this difficult nonlinear problem as a linear problem via the generator of the flow.
We use a low-order approximation method known as ``Ulam's method for the generator'':  we (indirectly) compute a time derivative of the matrix produced by the usual Ulam's method;  see \cite{FJK13}.
This approach, which is strongly related to the upwind scheme in finite volume methods \cite{leveque} has many advantages, perhaps most importantly the considerably reduced numerical effort as time integration of trajectories is no longer required as in the standard Ulam approach \cite{Ulam,DellnitzJunge99}.
Once in this linear (but high-dimensional) form, we use finite difference methods to create a linear program which may be solved using standard software (e.g. a standard implementation of simplex or barrier methods).
We illustrate our numerical approach on a steady two-dimensional flow and a periodically driven two-dimensional flow.

Our approach is very flexible:  it allows the trivial introduction of practical constraints, such as prohibiting perturbations (i) in certain parts of the spatial domain and (ii) at certain times in the case of periodic driving.
Moreover, it is also trivial to make the size of the perturbation depend on either space or time, or both.
Our results show that by making small changes to the vector field, dramatic changes in the dynamics and mixing of the fluid flow can be achieved.

The paper is arranged as follows.
In Section \ref{sec:generator} we describe in more detail our mathematical setting and define the generator for the steady and periodically driven situations.
Section \ref{sec:gennum} outlines the numerical approximation of the generator;  specifically Ulam's method for the generator \cite{FJK13}.
In Section \ref{sec:model} we develop the linear program which forms the core of the paper.
We begin by introducing the necessary constraints to ensure that the feasible region is the class of matrices that are interpretable as Ulam generators.
We then move on to the constraints that ensure that the perturbed Ulam generator corresponds to a generator for a vector field that is a small perturbation of the original vector field, and finally, we construct the objective itself.
Having performed the numerical optimisation, Section \ref{sec:interp} then describes how to extract a perturbed vector field from the perturbed Ulam generator matrix.  

Two case studies are presented in Section \ref{sec:numerics}.
The first is a steady two-dimensional flow.
 In this setting the advective dynamics is rather rigid:  particle trajectories must either move around closed curves or approach equilibrium points.
We show that even in this rather inflexible situation, one can still make significant improvements in the mixing rate by small modifications to the vector field.
The second case study is a periodically-driven two-dimensional flow where the advective dynamics displays both chaotic and regular (integrable) regimes.
We demonstrate how a relatively small perturbation can achieve a large increase in the size of the chaotic region and a corresponding decrease in the regular region, leading to a sharp improvement in mixing.
Finally, for the same system, we specifically target the regular regions and using small perturbations, (i) destroy the regular regions as much as possible, and (ii) enhance the regular regions as much as possible.


\section{Generators and their spectrum}
\label{sec:generator}

Our domain $M\subset \mathbb{R}^d$ is a compact, smooth manifold with vanishing\footnote{the obvious extensions can be made for general Riemannian manifolds.} curvature and $m$ denotes the volume measure on $M$.
We begin by discussing the autonomous setting before extending the discussion to the periodically-driven case.
We denote by $v:M\to \mathbb{R}^d$ the vector field that generates the flow;  we assume that $v$ is at least $C^1$ and that unique solutions of $\dot{x}=v(x)$ exist in both time directions.
The flow map generated by the vector field is denoted $\Phi:\mathbb{R}\times M\to M$, and $\Phi(t,x)\in M$ is the location of a trajectory after beginning at $x\in M$ and flowing for $t$ units of time.
We assume that $\Phi$ preserves an absolutely continuous invariant probability measure $\mu$;  that is, $\mu(\Phi(t,A))=\mu(A)$ for all Borel measurable $A\subset M$ and for all $t\in \mathbb{R}$.
For example, if the flow is incompressible (i.e.\ $v$ is divergence-free), then $\mu$ is normalised volume:  $\mu(\cdot)=m(\cdot)/m(M)$.
We denote the density of $\mu$ with respect to Lebesgue by $h=d\mu/dm$.
Associated with the flow map $\Phi$ is the transfer (or Perron-Frobenius) operator $\mathcal{P}:\mathbb{R}\times L^1(M,m)\to L^1(M,m)$ defined by $$\mathcal{P}(t,f)=f\circ\Phi(-t,x)\cdot|\det D\Phi(-t,x)|.$$
The operator $\mathcal{P}(t,\cdot)$ is the natural push-forward on densities for $t$ units of time, and invariance of $h$ can be stated as $\mathcal{P}(t,h)=h$ for all $t\in \mathbb{R}$.
The transfer operator is generated by the \emph{infinitesimal operator} or \emph{generator} $\mathcal{A}:D(\mathcal{A})\to L^1(M)$, defined by
\begin{equation}
\label{gen_limit}
\mathcal{A}f=\lim_{t\to 0}\frac{\mathcal{P}(t,f)-f}{t},
 \end{equation}
 where $D(\mathcal{A})$ is the set of $L^1$ functions for which the limit exists in the strong $L^1$ sense.
From this definition, it is clear that $\mathcal{A}h=0$.
It is well-known that for continuously differentiable $f$, one has
$$\mathcal{A}f=-\nabla\cdot (vf),$$
and that functions $u(t,x)=\mathcal{P}(t,u(x,0))$ satisfy the continuity equation
$$\frac{\partial u}{\partial t}+-\nabla\cdot (vu)=0.$$
The operator $\mathcal{A}$ ``generates'' $\mathcal{P}$ because $\mathcal{A}f=\lambda f\Leftrightarrow \mathcal{P}(t,f)=e^{\lambda t}f$ for $f\in L^1$, $\lambda\in \mathbb{C}$ and $t\ge 0$;  this is the content of the Spectral Mapping Theorem \cite{Pazy} (Theorem 2.2.4).

In the sequel we consider the generator of the SDE
\begin{equation}
\label{sde}dx_t=v(x_t)+\epsilon dw_t,
\end{equation}
 for small $\epsilon>0$, with reflecting boundary conditions, and where $dw_t$ represents Brownian motion.
The reason for the introduction of a Wiener process is mainly to simplify the functional analytic setup.
The corresponding generator is
\begin{equation}
\label{gen_eps}
\mathcal{A}f=-\nabla\cdot(vf)+\frac{\epsilon^2}{2}\triangle f,
\end{equation}
and one is concerned with solutions of the Fokker-Planck (or forward Kolmogorov or advection-diffusion) equation
$$\frac{\partial u}{\partial t}=-\nabla\cdot(vu)+\frac{\epsilon^2}{2}\triangle u,$$
initialised with some $u(0,\cdot)$, and with zero Neumann boundary conditions on $\partial M$ (the latter ensure that mass is preserved in $M$ by this ``no flux'' condition).
The operator $\mathcal{A}$ is elliptic, but not self-adjoint in general, and one can find a $c<0$ such that $\sigma(\mathcal{A})\cap\{z\in \mathbb{C}: \Re z\ge c\}$ consists only of the eigenvalue $\lambda_1=0$.
Associated with the eigenvalue $\lambda_1=0$ is the unique equilibrium concentration $h$ of the associated stochastic process:  $\mathcal{A}h=0$.
The real part of the eigenvalue $\lambda_2$ closest to the imaginary axis controls the mixing rate of the SDE (more negative, faster mixing), and it is this spectral gap between $0$ and $\Re\lambda_2$ that we wish to widen through perturbation of $v$.

\subsection{Periodically driven flows}
\label{sect:periodic}

For periodically-driven flows we have $v:M\times \tau S^1\to \mathbb{R}^d$, where $\tau$ is the driving period and $S^1$ is the circle of unit circumference.
The flow map is $\Phi:\mathbb{R}\times \tau S^1\times M\to M$, and $\Phi(t,s,x)\in M$ is the location of a trajectory after beginning at $x\in M$ at a time $s$, and flowing for $t$ units of time.
Associated with the flow map $\Phi$ is the transfer (or Perron-Frobenius) operator $\mathcal{P}:\mathbb{R}\times \tau S^1\times L^1(M,m)\to L^1(M,m)$ defined by $$\mathcal{P}(t,s,f)=f\circ\Phi(-t,s,x)\cdot|\det D\Phi(-t,s,x)|.$$
If we were to take the limit (\ref{gen_limit}) we would obtain a different operator $\mathcal{A}_s$ at each $s\in \tau S^1$, and none of these would generate $\mathcal{P}$ in the usual sense.
Therefore, following the constructions in \cite{FrKo} we augment the phase space by defining $\bm{M}=\tau S^1\times M$.
Further, as in the autonomous setting, we add a Brownian motion term to yield a simpler spectral theory.
We have
\begin{eqnarray*}
d\theta_t & = & dt,\\
dx_t & = & v(\theta_t,x_t)dt + \epsilon dw_t,
\end{eqnarray*}
or alternatively,
\begin{equation}
d\bm{x}_t = \bm{v}(\bm{x}_t)dt + \mathbb{\epsilon}d\bm{w}_t
\label{eq:augSDE}
\end{equation}
where the augmented state~$\bm{x}=(\theta,x)$, the vector field~$\bm{v}=(1,v(\theta,x))$, and $\{\bm{w}_t\}$ is a~$(d+1)$-dimensional standard Wiener process, and
\[
\bm{\epsilon} = \begin{pmatrix}
\epsilon I_{d\times d} & 0\\
0 & 0
\end{pmatrix}\in\mathbb{R}^{(d+1)\times(d+1)},
\]
where~$I_{d\times d}\in\mathbb{R}^{d\times d}$ is the identity matrix.
The generator on this augmented phase space can be obtained from the Fokker--Planck equation associated with~\eqref{eq:augSDE}.
Let $\bm{f}:\bm{M}\to \mathbb{R}$ and denote $f_{\theta}$ on each time fibre $\theta$ by $f_\theta=\bm{f}(\theta,\cdot)$.
We have
\begin{equation}
\bm{A} \bm{f}(\theta,x) = -\partial_{\theta}\bm{f}(\theta,x) + \mathcal{A}_{\theta}f_{\theta}(x),
\label{eq:augIG}
\end{equation}
where by (\ref{gen_eps}), one has $$\mathcal{A}_\theta f=-\nabla\cdot(v(\theta,\cdot)f)+\frac{\epsilon^2}{2}\triangle f.$$
We call~$\bm{A}$ the \emph{augmented generator}.
The spectral structure of $\bm{A}$ in the periodically driven setting is similar to the spectrum of $\mathcal{A}$ in the autonomous setting;  see \cite{FrKo} for details.
In particular, we wish to widen the spectral gap between the imaginary axis and the eigenvalue $\lambda_2$ with real part closest to the imaginary axis.


\section{Numerical approximation of the generator}
\label{sec:gennum}

Our strategy for optimally perturbing the vector field $v$ in some pre-specified neighborhood to maximize the rate of mixing is to work directly with the generator rather than the vector field.
In order to practically optimize the generator we require some finite-dimensional numerical representation.
Moreover, it is advantageous if there is a simple connection between this numerical representation and the underlying vector field.
Our numerical method of choice is the ``Ulam's method for the generator'' method, developed in \cite{FJK13}, and we refer the reader to this paper for details beyond those presented here.
The recent work of \cite{FrKo} extends this methodology to the situation of periodically driven dynamics, which is of particular interest in fluid dynamics.
We describe the method first for autonomous vector fields and then the required modifications for periodically-driven systems.

Partition $M$ into rectangular sets of identical size to form $\mathcal{B}_n=\{B_1,\ldots,B_n\}$, and let $\Xi_n$ denote the span of the indicator functions on the elements of $\mathcal{B}_n$.
The standard Ulam's method \cite{Ulam} to estimate the action of transfer operators on $\Xi_n$  can be adapted to create an approximate generator $\mathcal{A}_n$ (also acting on $\Xi_n$) by using (\ref{gen_limit}) with $\mathcal{P}$ replaced by its Ulam version;  see \cite{FJK13}.
It may then be shown that the matrix representation of $\mathcal{A}_n:\Xi_n\circlearrowleft$, under left multiplication, is given by the following formula:
\begin{equation}\label{ulamformula2}
 (A_n)_{ij}=\left\{
            \begin{array}{ll}
              \displaystyle \frac{1}{m(B_j)}\int_{B_i\cap B_j}\max\{v(x)\cdot \mathbf{n}_{ij},0\}\ dm_{d-1}(x), & \hbox{$i\neq j$;} \\
              \displaystyle -\sum_{j\neq i} \frac{m(B_j)}{m(B_i)}(A_n)_{ij}, & \hbox{otherwise.}
            \end{array}
          \right.,
\end{equation}
where $m_{d-1}$ is the $d-1$-dimensional volume measure induced on co-dimension 1 surfaces (used here to integrate over a face of a rectangular prism), and $\mathbf{n}_{ij}$ is the unit normal vector pointing out of $B_i$ into $B_j$ (or the zero vector if $B_i\cap B_j$ is empty).
Given $f=\sum_{i=1}^n f_i\mathbf{1}_{B_i}\in \Xi_n$, one computes $\mathcal{A}_n(f)=\sum_{j=1}^n\left(\sum_{i=1}^n (A_n)_{ij}f_i\right)\mathbf{1}_{B_j}$.
It is shown in Lemma 4.7 \cite{FJK13} that there exists a nonnegative, nonzero $h_n\in\Xi_n$ with $\mathcal{A}_n(h_n)=h_n$.
Moreover, the spectrum of $A_n$ is confined to the left half of the complex plane.
The matrix $A_n$ is related to the upwind scheme in finite-volume methods \cite{leveque} and the spatial discretisation of $M$ into $\mathcal{B}_n$ induces a numerical diffusion.
Thus, as remarked in \cite{FJK13}, the generator $\mathcal{A}_n$ is in fact a better numerical approximation of an SDE of the type (\ref{sde}) for suitable $\epsilon$ and covariance matrix, than the deterministic ODE $\dot{x}=v(x)$.
In fact, it is shown in \cite{FJK13} that $A_n$ can be regarded as a rate matrix for a Markov jump process, identifying each box as a state.
In our computations, we use a partition $\mathcal{B}_n$ into cubes\footnote{One can alternatively efficiently include an additional discrete isotropic diffusion term as in \cite{FrKo} to achieve isotropic diffusion with a rectangular grid.}, or approximate cubes, so as to make this numerical diffusion as isotropic as possible.
We therefore think of $\mathcal{A}_n$ as an approximate generator for a process of the type (\ref{sde}).
We compute the integral in (\ref{ulamformula2}) by numerical quadrature.


\subsection{Periodically driven case}

In the periodically-driven setting, we use exactly the same constructions as above, but now $\mathcal{B}_n$ is a partition of the augmented space $\bm{M}$ into rectangles with are approximately cubic.
We remark that if two rectangles $B_i$ and $B_j$ intersect at a face perpendicular to the time direction, the integral in (\ref{ulamformula2}) is simply $1/\mbox{(diameter of the set $B_j$ in the time direction)}$;  in particular, no quadrature is required for these faces.
%
%

\section{A linear program to enhance mixing}
\label{sec:model}

Recall that we wish to increase the $L^2$-mixing rate $\Lambda$ (defined in (\ref{L2mix})) of the SDE (\ref{sde}) (resp.\ (\ref{eq:augSDE})) in the autonomous (resp.\ periodically-driven) setting, by increasing the spectral gap of $A_n$ from the imaginary axis.
%
We now discuss this problem of optimising the spectrum of $A_n$ subject to necessary and desirable constraints.
First, we attend to the necessary constraints.

\subsection{Constraints to ensure we have a valid Ulam generator}

Suppose that we perturb the matrix $A_n$ by adding a perturbation matrix $E$.
The resulting matrix $A_n+E$ should satisfy various conditions.
Note that all of the following constraints are linear in $E$.

\paragraph{Conservation of mass:}  Conservation of mass is expressed as $\mathcal{A}_n^*(\mathbf{1})=A_n\mathbf{1}=0$, where $\mathcal{A}_n^*$ is the $L^2$-adjoint of $\mathcal{A}_n$.  We require that the perturbed generator also conserve mass.
In terms of $A_n$ this is expressed as $\mathbf{1}A_n^\top=0$.
Thus we insist that
$$\sum_{j=1}^n E_{ij}=0\mbox{ for all }1\le i\le n.$$

\paragraph{Signs of entries of $A_n+E$:}  A necessary condition for $A_n+E$ to be a genuine generator is that the signs of the entries of $A_n+E$ match those prescribed in (\ref{ulamformula2}).
In particular, diagonal entries are non-positive and off-diagonal entries are non-negative, or in symbols, $$A_{n,ij}+E_{ij}\ge 0\mbox{ for }i\neq j,\mbox{ and }A_{n,ii}+E_{ii}\le 0\mbox{ for }1\le i\le n.$$

\paragraph{Preservation of the same equilibrium distribution:}  Suppose that $h_n$ is the invariant density of the approximate generator $\mathcal{A}_n$; one has $\mathcal{A}_n(h_n)=0$, or $h_nA_n=0$.  After adding the perturbation, it is likely that we wish to preserve the same invariant density.
An important example of this principle is that a volume-preserving flow should remain volume-preserving after the perturbation.
Thus, we require
$$\sum_{i=1}^n h_{n,i}E_{ij}=0\mbox{ for each } 1\le j\le n,$$ where $h_{n,i}$ are the coefficients of the step function $h_n=\sum_{i=1}^n h_{n,i}\mathbf{1}_{B_i}$.

\paragraph{Flow can only occur through rectangles with adjacent faces:} If two rectangles $B_i$ and $B_j$ do not meet at a face then there can be no flow directly from $B_i$ to $B_j$.  Thus, $$A_{n,ij}+E_{ij}=0,\mbox{ if $B_i$ and $B_j$ do not meet at a face}.$$
  In fact, at the discrete level of rectangles, there should not be flow out of $B_i$ into $B_j$ \emph{and} vice-versa.  A simple way to enforce this, which is only a mild restriction on the generality of $E$, is to allow outflow from $B_i$ into $B_j$ according to the perturbed generator only when there was already such an outflow for the original generator.
  We do this in all reported numerical experiments in Section \ref{sec:numerics}.

\subsection{Constraints to ensure the perturbation is small}

At this point, if $A_n+E$ satisfies the above conditions, it is a perfectly good Ulam generator.
However, by making the entries of $E$ larger, which corresponds to making the underlying flow faster, we can trivially speed up the rate of mixing simply by putting e.g.\ $E=\beta A_n$ for some large $\beta>0$;  this would correspond to replacing $v$ with $(1+\beta) v$.
Our goal is to maximize the mixing rate while only perturbing within a \emph{small neighborhood} of the original vector field.
Thus, we require additional constraints on $E$ to remain in such a neighborhood.
The discussion below is for the autonomous setting;  afterwards we describe the necessary changes in the periodically driven settting.
The user may be satisfied with only a subset of the following constraints, for example, only constraining the optimized velocity field or the optimal perturbation, but not both.
For completeness we describe how to constrain both.

By inspecting (\ref{ulamformula2}), we see that we may interpret $A_{n,ij}$ as the mean, normal, outward velocity on the face $B_i\cap B_j$ divided by the diameter of the rectangles in this normal direction.
Therefore, given a matrix $A_n$, we can infer a corresponding spatially discretized vector field at the centroids of the rectangle faces, denoted $v^+_{n,ij}$.
Explicitly,
\begin{equation}
\label{velconvert}
v^+_{n,ij}=A_{n,ij}\cdot \Delta_{ij},
 \end{equation}
 where $\Delta_{ij}$ is the rectangle diameter in the direction normal to $B_i\cap B_j$.
As we assume that the rectangles comprising $\mathcal{B}_n$ are identical, there are only $d$ distinct values for the $\Delta_{ij}$.
We now detail several constraints to control the perturbed, discrete vector field, denoted by $\tilde{v}^+_{n,ij}$.
We denote by $\tilde{v}$ a vector field created from $\tilde{v}^+_{n,ij}$ by a smooth interpolation.

In the time-periodic case, we apply (\ref{maxspeed})--(\ref{maxderivpert}) only when $B_i\cap B_j$ has normal pointing in a spatial (not the temporal) direction.
If $B_i\cap B_j$ has normal in the time direction we set $E_{ij}=0$ as we cannot adjust the flow of time.

\paragraph{No increase in the maximum speed of the perturbed velocity field:}  We wish to ensure that $\|\tilde{v}\|_\infty\le \|v\|_\infty$\footnote{One could easily allow a small increase if one wishes.}.
Suppose that $B_i$ and $B_j$ are boxes sharing a face. 
At the discrete level, we want to ensure that $\max_{1\le i,j\le n}|v_{n,ij}|$ is not increased by the perturbation.
To achieve this we set
\begin{equation}
\label{maxspeed}
A_{n,ij}+E_{ij}\le \max_{1\le i\neq j\le n}A_{n,ij}.
\end{equation}

\paragraph{No increase in the mean speed of the perturbed velocity field:} The rate of total flux exiting a rectangle $B_i$ is captured by the diagonal element $A_{n,ii}$.  We can therefore ensure that the rate of total flux over all of the phase space $X$ is not increasing by insisting that
\begin{equation}
\label{globalspeed}
\sum_{i=1}^n E_{ii}\ge 0,
\end{equation}
noting that the diagonal elements of $A_n$ are non-positive.

\paragraph{Locally limiting the velocity perturbation:}  We wish to remain in a $C^0$ neighborhood of the original velocity field, i.e.\ $\|\tilde{v}-v\|_\infty$.
Suppose that $B_i$ and $B_j$ are boxes sharing a face. 
At the discrete level, we desire $|\tilde{v}^+_{n,ij}-v^+_{n,ij}|=|E_{ij}|\Delta_{ij}\le \epsilon_1$, by (\ref{velconvert}) where $\epsilon_1$ is our allowed $C^0$ deviation.
We therefore constrain
\begin{equation}
\label{maxpert}
|E_{ij}|\le \epsilon_1/\Delta_{ij},\qquad 1\le i\neq j\le n.
\end{equation}

\begin{figure}[hbt]
  \centering
  \includegraphics[width=0.5\textwidth]{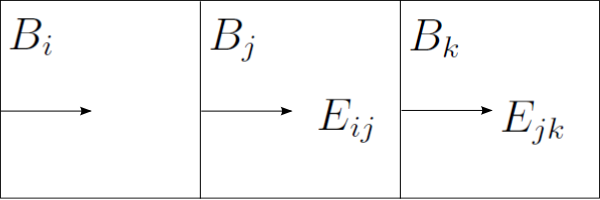}\\
  \caption{Schematic for the approximation in (\ref{maxderiv}).}\label{fig:Box1}
\end{figure}

\begin{figure}[hbt]
  \centering
  \includegraphics[width=0.3\textwidth]{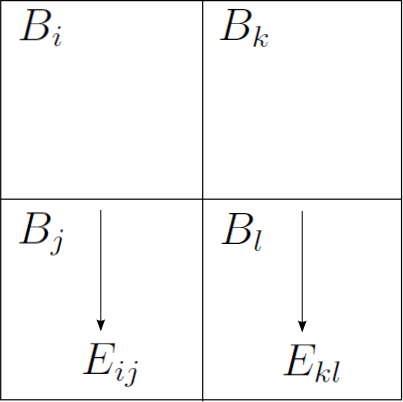}\\
  \caption{Schematic for the approximation in (\ref{maxderiv2}).}\label{fig:Box2}
\end{figure}

\paragraph{Locally limiting the maximum spatial derivative of the perturbed velocity field:}  High rates of mixing can be achieved by spatially irregular vector fields (see for example \cite{LinThiffeaultDoering}).
Using the discrete generator $A_n$, we can estimate the derivatives of the perturbed vector field by taking differences of the discrete velocity estimates in (\ref{maxspeed}) in each of the coordinate directions.
For example in two dimensions,
\begin{eqnarray}
\label{maxderiv}
|\tilde{v}^+_{ij}-\tilde{v}^+_{jk}|&=&|(A_{n,ij}+E_{ij})-(A_{n,jk}+E_{jk})|\Delta_{ij}\le \epsilon_2,\\
\label{maxderiv2}
|\tilde{v}^+_{ij}-\tilde{v}^+_{kl}|&=&|(A_{n,ij}+E_{ij})-(A_{n,kl}+E_{kl})|\Delta_{ij}\le \epsilon_2,
\end{eqnarray}
where $\epsilon_2$ is our allowed spatial derivative bound and in (\ref{maxderiv}) the faces $B_i\cap B_j$ and $B_j\cap B_k$ are opposing (Figure \ref{fig:Box1}), while in (\ref{maxderiv2}) the faces $B_i\cap B_j$ and $B_k\cap B_l$ are adjacent (Figure~\ref{fig:Box2}).


\paragraph{Locally limiting the maximum spatial derivative of the perturbation:}
Our aim is to make a small $C^1$ perturbation of the original vector field.
The spatial derivative of the perturbation can be estimated and constrained by:
\begin{eqnarray}
\label{maxderivpert}
|E_{ij}-E_{jk}|&\le& \epsilon_3/\Delta_{ij},\\
\label{maxderivpert2}
|E_{ij}-E_{kl}|&\le& \epsilon_3/\Delta_{ij},
\end{eqnarray}
where $\epsilon_3$ is our allowed spatial derivative bound, and in (\ref{maxderivpert}) the faces $B_i\cap B_j$ and $B_j\cap B_k$ are opposing (Figure \ref{fig:Box1}), while in (\ref{maxderivpert2}) the faces $B_i\cap B_j$ and $B_k\cap B_l$ are adjacent (Figure \ref{fig:Box2}).

\subsection{The objective}
Consider a family of matrices $A_n+t{E}$, where ${E}$ is a small-magnitude perturbation matrix and $t\in \mathbb{R}$.
Because there is no simple expression for the eigenvalues of $A_n+t{E}$ in terms of the eigenvalues of $A_n$, we will use a linearised estimate for the eigenvalues of $A_n+t{E}$ based on the derivatives of the eigenvalues of $A_n$ with respect to $t$.
Such an estimate will be valid because $E$ is small; we will evaluate this linearised estimate at $t=1$.
By standard arguments (e.g. Theorem 6.3.12~\cite{hornjohnson}), one has
\begin{equation}
\label{star}
\frac{d\lambda_k}{dt}=(y^k)^*{E}w^k,
 \end{equation}
 where $\lambda_k\in\mathbb{C}$ is the $k^{th}$ eigenvalue of $A_n$, and $y^k$ (resp.\ $w^k$) are the left (resp.\ right) eigenvectors of $A_n$, normalised so that $(w^k)^*w^k=1$ and $(y^k)^*w^k=1$, $k=1,\ldots,K$.
We are interested in the real part of $\lambda_k$;  from (\ref{star}) one easily computes
$$\frac{d}{dt}\left(\Re\lambda_k\right)=\Re(y^k)^\top E\Re(w^k)+\Im(y^k)^\top E\Im(w^k).$$

\subsection{The full linear program}
\label{lpsection}

Below we write a simple linear program to select an $E$ to push the eigenvalues $\lambda_2,\ldots,\lambda_K$ away from the imaginary axis.
Note that $\lambda_k,\Re(w^k),\Im(w^k),\Re(y^k),\Im(y^k)$, $k=2,\ldots,K$ are all known constants, precomputed from $A_n$, and that $\epsilon_1, \epsilon_2, \epsilon_3, \Delta_{ij}$ are fixed constants.
The only variables are the entries of $E$ and the auxiliary variable $z\in\mathbb{R}$.
\begin{eqnarray}
\label{objective}\min_{E_{ij}}&&z\\
\label{multiobj}\mbox{s.t.}&&z\ge \lambda_k+\sum_{i,j=1}^n\left(\Re(y^k_i)^\top\Re(w^k_j)+\Im(y^k_i)\Im(w^k_j)\right)E_{ij},\quad k=2,\ldots,K\\
\label{stochasticity}&&\sum_{j=1}^n E_{ij}=0, \quad 1\le i\le n \\
\label{invariance}&&\sum_{i=1}^n y^1_iE_{ij}=0,\quad 1\le j\le n  \\
\label{diagonal}&&A_{ii}+E_{ii}\le 0,\quad 1\le i\le n\\
\label{offdiagonal}&&A_{ij}+E_{ij}\ge 0, \quad 1\le i\neq j\le n\\
\label{locality}&& E_{ij}=0,\mbox{ if $B_i\cap B_j=\emptyset$\quad }\\
\label{maxspeed2}&& A_{n,ij}+E_{ij}\le \max_{1\le i\neq j\le n}A_{n,ij}\\
\label{globalspeed2}&& \sum_{i=1}^n E_{ii}\ge 0\\
\label{casemaxpert}
&&\left.
\begin{array}{ll}
E_{ij}\le \epsilon_1/\Delta_{ij}&\\
E_{ij}\ge -\epsilon_1/\Delta_{ij}
\end{array}
\right\}\\
\label{casemaxderiv}
&&\left.
\begin{array}{ll}
(A_{ij}+E_{ij})-(A_{jk}+E_{jk})\le\epsilon_2/\Delta_{ij}&\\
(A_{ij}+E_{ij})-(A_{jk}+E_{jk})\ge-\epsilon_2/\Delta_{ij}
\end{array}
\right\}\quad\mbox{if $B_i\cap B_j$ is opposing $B_j\cap B_k$}\\
\label{casemaxderiv2}
&&\left.
\begin{array}{ll}
(A_{ij}+E_{ij})-(A_{jk}+E_{jk})\le\epsilon_2/\Delta_{kl}&\\
(A_{ij}+E_{ij})-(A_{jk}+E_{jk})\ge-\epsilon_2/\Delta_{kl}
\end{array}
\right\}\quad\mbox{if $B_i\cap B_j$ is adjacent to $B_k\cap B_l$}\\
\label{casemaxderivpert}
&&\left.
  \begin{array}{ll}
    E_{ij}-E_{jk}\le\epsilon_3/\Delta_{ij} &  \\
    E_{ij}-E_{jk}\ge-\epsilon_3/\Delta_{ij}
  \end{array}
\right\}\quad\mbox{if $B_i\cap B_j$ is opposing $B_j\cap B_k$}\\
\label{casemaxderivpert2}
&&\left.
  \begin{array}{ll}
    E_{ij}-E_{kl}\le\epsilon_3/\Delta_{ij} &  \\
    E_{ij}-E_{kl}\ge-\epsilon_3/\Delta_{ij}
  \end{array}
\right\}\quad\mbox{if $B_i\cap B_j$ is adjacent to $B_k\cap B_l$}\\
\end{eqnarray}
The objective is to select a generator perturbation $E$ so that the linearised estimates of the $\lambda_k$, $k=2,\ldots,K$ after perturbation by $E$ are all as far from the imaginary axis as possible, subject to the constraints.
The constraints (\ref{stochasticity}) enforce mass conservation.
The constraint (\ref{invariance}) ensures that the same invariant density is preserved before and after perturbation.
The constraints (\ref{diagonal})--(\ref{locality}) ensure that $A+E$ is a valid Ulam generator matrix (correct signs of diagonal and off diagonal terms, and flowing only to neighbouring boxes).

The constraint (\ref{maxspeed2}) limits the velocity of the perturbed velocity field as per (\ref{maxspeed}).
The inequality (\ref{globalspeed2}) ensures no increase in total flux as per (\ref{globalspeed}).
The constraints (\ref{casemaxpert}) limit the size of the perturbation as per (\ref{maxpert}).
The constraints (\ref{casemaxderiv})--(\ref{casemaxderiv2}) and (\ref{casemaxderivpert})--(\ref{casemaxderivpert2}) limit the spatial deriviative of the velocity field and the perturbation, respectively, as per the constraints (\ref{maxderiv})--(\ref{maxderiv2}) and (\ref{maxderivpert})--(\ref{maxderivpert2}), respectively.

Solving for the optimal $E$ is fast, even for large $n$  (there are only about $2n$ variables in $E$ for 2D flows).
Once we have an optimal $E$, we can infer a perturbed vector field $\tilde{v}$ corresponding to the Ulam generator $A_n+E$ as described in the next section.
\paragraph{Further practical constraints and generalisations:}
Our approach is flexible and practical;  for example, if one can only alter the vector field physically at certain spatial locations, then it is easy to add constraints like $E_{ij}=0$ at boxes $B_i, B_j$ where one cannot physically alter the vector field.
Similarly, in the time-periodic case one can also enforce spatial perturbations only at certain times in the driving cycle.
This can be achieved by setting $E_{ij}=0$ where $B_i\cap B_j$ is a face normal to the time direction corresponding to a time at which perturbation is not possible.
If one wishes the perturbation size controlled by $\epsilon_1, \epsilon_2, \epsilon_3$ to depend on space or time, these constants can depend on $i$ and $j$ with no change to the complexity of the linear program.



\section{Inferring a vector field from the generator}
\label{sec:interp}

Equation (\ref{ulamformula2}) describes how to construct $\mathcal{A}_n$ from a vector field $v$.
We now discuss the reverse operation, focussing on the case of divergence-free flows, corresponding to incompressible fluids.

\subsection{Linear interpolation of the vector field}

Let $\{v(x)\}_{x\in M}$ be a steady vector field;  we describe the minor modifications required in the time-periodic case at the end of this section.

\begin{figure}[hbt]
  \centering
  \includegraphics[width=0.7\textwidth]{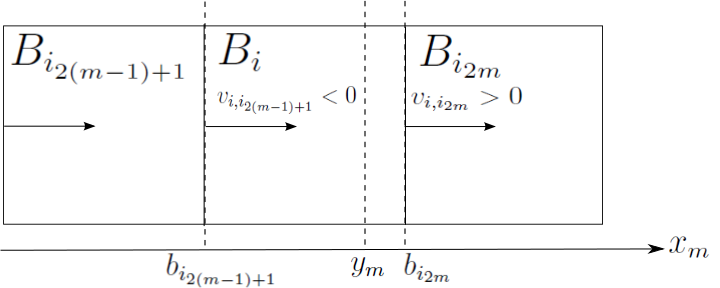}\\
  \caption{Interpolation to determine the velocity in the direction of $x_m$.}\label{fig:interp}
\end{figure}

Suppose that we have partitioned $M$ into cubes $\mathcal{B}_n$, aligned with the standard coordinate directions in $\mathbb{R}^{d}$.
Consider a fixed rectangle $B_i\in\mathcal{B}_n$, and its at most $2d$ neighbours $B_{i_1},\ldots,B_{i_{2d}}$, some of which may be empty.
We locally index these neighbouring boxes so that for $m=1,\ldots,2d$, boxes $B_{i_{2(m-1)+1}}$ and $B_{i_{2m}}$ correspond to intersections with $B_i$ on faces with normals in the coordinate directions $-x_m$ and $x_m$, respectively; see Figure \ref{fig:interp}.

By (\ref{ulamformula2}) the value $A_{i,i_k}\Delta_{i,i_k},$ is the average outward-pointing velocity on the shared face $B_i\cap B_{i_k}$, pointing out of $B_i$ into $B_{i_k}$, $k=1,\ldots,2d$.
Define a signed mean velocity on the face $B_i\cap B_{i_k}$ (positive velocity is outflow, negative velocity is inflow):
\begin{equation}
\label{signedvel}
v_{i,i_k}=\left\{
  \begin{array}{ll}
    A_{i,i_k}\Delta_{i,i_k}, & \hbox{if $A_{i,i_k}>0$;} \\
    -A_{i_k,i}\Delta_{i,i_k}, & \hbox{otherwise.}
  \end{array}
\right.
\end{equation}
By divergence-freeness one has
\begin{equation}
\label{divfree}
\sum_{k=1}^{2d} \frac{m(B_i)}{\Delta_{i,i_k}}v_{i,i_k}=0;
\end{equation}
in words, the rate of total outflow from $B_i$ equals the rate of total inflow to $B_i$.
We are now ready to define an interpolated divergence-free vector field $\tilde{v}(x)$.
Consider a point $(y_1,\ldots,y_d)\in B_i$, and denote by $b_{i_{2(m-1)+1}}$ and $b_{i_{2m}}$ the $x_m$ coordinates of the faces $B_i\cap B_{i_{2(m-1)+1}}$ and $B_i\cap B_{i_{2m}}$;  see Figure \ref{fig:interp}.
We define the $m^{th}$ coordinate of $\tilde{v}(y_1,\ldots,y_d)$, corresponding to the coordinate direction, to be
\begin{equation}
\label{interp}
[\tilde{v}(x)]_m:=-v_{i,i_{2(m-1)+1}}\frac{(y_m-b_{i_{2(m-1)+1}})}{(b_{i_{2m}}-b_{i_{2(m-1)+1}})}+ v_{i,i_{2m}}\frac{(b_{i_{2m}}-y_m)}{(b_{i_{2m}}-b_{i_{2(m-1)+1}})}.
\end{equation}
This is simply a convex combination (a linear interpolation) of the velocities on the opposing faces normal to the $m^{\rm th}$ coordinate direction $x_m$.
The negative sign of $v_{i,i_{2(m-1)+1}}$ appears because this velocity is relative to the $-x_m$ coordinate direction.
Differentiating with respect to $x_m$, replacing $b_{i_{2m}}-b_{i_{2(m-1)+1}}$ with $\Delta_{i,{2m}}$ (possible because $B_i$ is a rectangle), and summing over $m=2,\ldots,d$, we see from (\ref{divfree}) (noting that $m(B_i)$ is constant in the sum) that $\tilde{v}$ is everywhere divergence-free.
%

Finally, we consider the situation where $v=v(t,x), t\in \tau S^1, x\in M$.
Our partition $\mathcal{B}_n$ is now a partition of the augmented space $\bm{M}$ into $d+1$-dimensional rectangles.
The interpolation proceeds exactly as above, except that since the velocity is constant in the time direction, no interpolation is necessary along this coordinate direction in augmented space.
Divergence-freeness of $v(t,\cdot)$ for each $t\in \tau S^1$ also follows as above by an identical argument.

\subsection{Construction and interpolation of a stream function when $\dim M=2$}
\label{sec:smooth}

The affinely interpolated vector field $\tilde{v}$ constructed in the previous section is piecewise continuous.
If $M$ is two-dimensional, one way to produce a smooth version of $\tilde{v}$ that continues to preserve area is to explicitly compute the stream function \cite{Batchelorbook} $\psi$ for $\tilde{v}$ and then smooth $\psi$ using one's favourite technique.
If the domain is itself a rectangle $M=[a,b]\times[c,d]$ aligned with the standard coordinate directions $(x_1,x_2)$, the computation of $\psi$ can be achieved in the following way.
We describe the case of steady $\tilde{v}$;  one can make obvious modifications for time-periodic $\tilde{v}$.
On the lower boundary face $[a,b]\times\{c\}$ of $M$, select a value for $\psi$ (e.g.\ without loss, set $\psi=0$ on this boundary face).
Using the fact that $\partial\psi/\partial y=[\tilde{v}]_1$ (where $[\tilde{v}]_1$ denotes the first coordinate of $\tilde{v}$), we define a grid of $x_1$-coordinates $\{z_1,\ldots,z_q\}\subset [a,b]$ on this face and then for each $z_r\in \{z_1,\ldots,z_q\}$ and all $y\in[c,d]$ set $\psi(z_r,y)=\int_{c}^{y}[\tilde{v}(z_r,s)]_1\ ds$.
As $\tilde{v}$ is piecewise affine, this integration can be done exactly.
One may now create a smooth version $\bar{\psi}$ of $\psi$ by interpolation and then obtain a smooth area-preserving vector field $\bar{v}$ as $\bar{v}(x,y)=(\partial \bar{\psi}/\partial y,-\partial\bar{\psi}/\partial x)$~\cite{Batchelorbook}. 

\section{Numerical results}
\label{sec:numerics}

\subsection{Case study 1: Single gyre}

\begin{figure}[H]
  \centering
  \includegraphics[width=7cm]{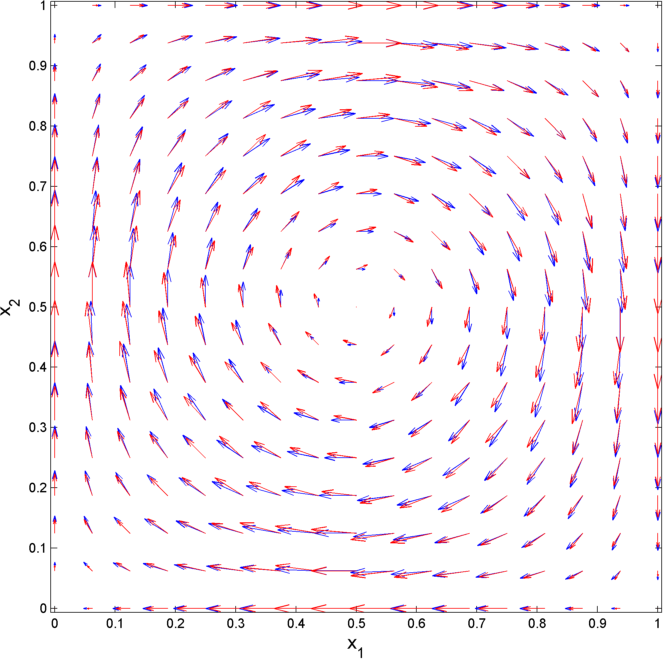}\\
  \caption{Comparison between the original (blue) and optimized (red) velocities for the rotating flow \eqref{singlegyreeqn}.}\label{quivers}
\end{figure}

\begin{figure}[H]
  \centering
  \includegraphics[scale=0.45]{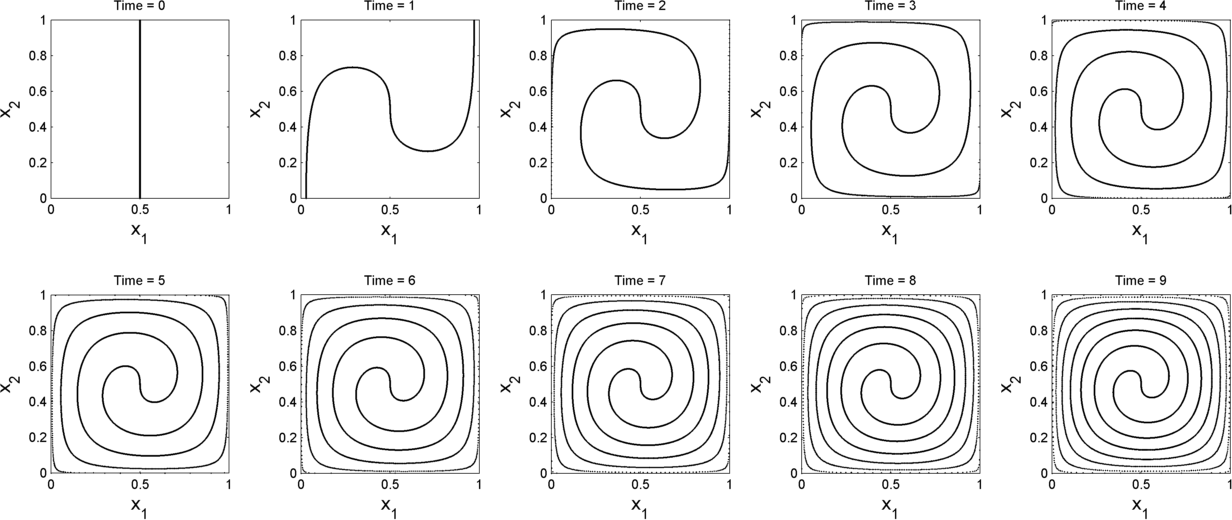}\\
  \caption{The evolution of a vertical line segment under the flow \eqref{singlegyreeqn} from time $T=0$ to $T=9$.}\label{line_evo}
\end{figure}

We begin by studying a simple rotating flow on the unit square $M=[0,1]^2$.
Let
\begin{equation}
\label{singlegyreeqn}
v(x)=(-\sin(\pi x_1)\cos(\pi x_2),\cos(\pi x_1)\sin(\pi x_2)),
\end{equation}
shown in blue in Figure \ref{quivers}.
One can easily verify that the $2\times 2$ matrix of spatial derivatives $Dv(x)$ is skew-symmetric with zero trace.
Thus, the flow is area-preserving, and the instantaneous local behaviour at every point is a pure rigid rotation;  that is, there is zero divergence and zero shear.
The domain $M$ is foliated by concentric closed curves and trajectories of the flow $\dot{x}=v(x)$ simply move around the curve (or streamline) they begin on.
Thus there is no mixing between trajectories.
The flow is not mixing, nor even ergodic, in the sense of ergodic theory;  each closed curve supports its own ergodic, but not mixing, dynamics.
It is not possible to improve this situation through perturbation.
Topological considerations (the Poincar\'e-Bendixson Theorem -- see e.g.\ \cite{katok}) state that all recurrent trajectories must be periodic (i.e.\ closed curves).
However, we note that evolution of curves \emph{transverse} to the streamlines display mixing-like behaviour; see Figure \ref{line_evo}.
Moreover, in the presence of small diffusion, global mixing occurs and one achieves a strictly negative $\Lambda$ in (\ref{Lambdaeqn}).

\begin{figure}[hbt]
  \centering
  \includegraphics[width=9cm]{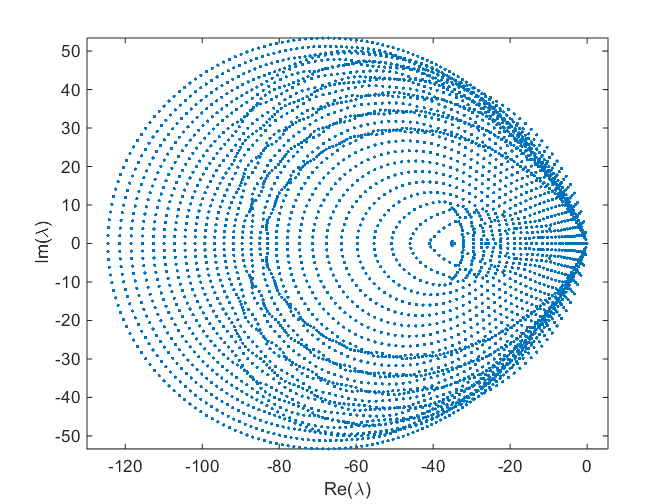}\\
  \caption{Spectrum of $A_n$ for the single gyre (\ref{singlegyreeqn}), $n=4096$.}\label{single_gyre_evals}
\end{figure}

We partition the unit square into a grid of $64\times 64$ identical squares and use (\ref{ulamformula2}) to create a $4096\times 4096$ matrix $A_n$ approximating the generator $\mathcal{A}$.
The spectrum of $A_n$ is shown in Figure \ref{single_gyre_evals}.
Note that the only spectral value intersecting the imaginary axis is the eigenvalue 0.

\begin{figure}[H]
  \centering
  \includegraphics[width=10cm]{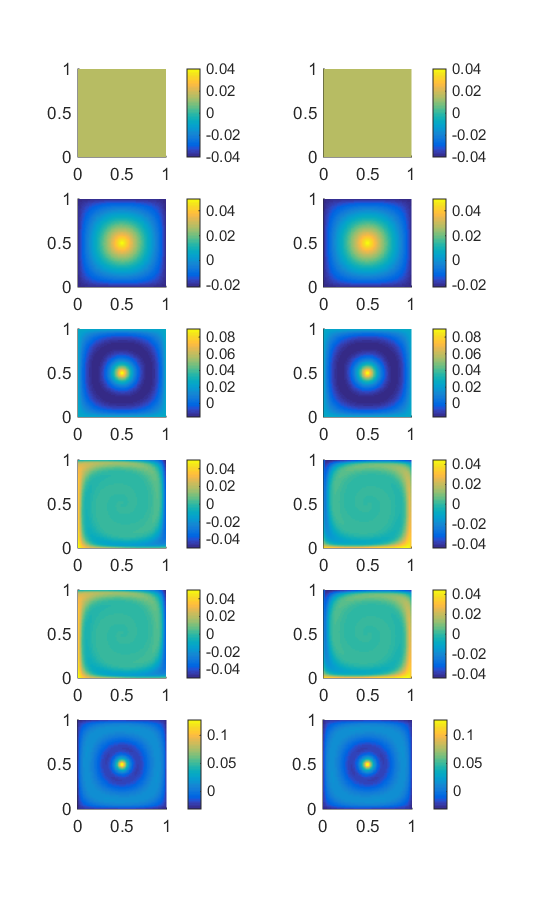}\\
  \caption{Real parts of eigenvectors of $A_n$ corresponding to the eigenvalues $\lambda_1,\ldots,\lambda_6$ in descending order of real part.  Upper: $\lambda_1$, lower: $\lambda_6$.  The left column shows left eigenvectors and the right column shows right eigenvectors.}\label{single_gyre_evecs}
\end{figure}

The real parts of the leading\footnote{those with real parts closest to the imaginary axis.} six left (resp.\ right) eigenvectors are shown in the left (resp.\ right) column of Figure \ref{single_gyre_evecs}.

The most important eigenfunctions from a dynamical point of view are those corresponding to $\lambda_1=0$ and $\lambda_2<0$.
We see from Figure \ref{single_gyre_evecs} that the eigenfunction corresponding to $\lambda_1$ is constant, indicating that area is preserved.
The second eigenfunction (2nd row of Figure \ref{single_gyre_evecs}) has a peak at $x=(1/2,1/2)$ and minima at the boundary of $M$.
This indicates that in the presence of small diffusion, it takes the longest time to transport measure from the centre of the gyre to its extremities.
In other words, the centre and the boundary are the most dynamically disconnected regions.

We compute $\max_{1\le i\neq j\le n}A_{n,ij}\approx 63.9743$ and $\Delta=1/64=0.015625$.
We choose parameters $\epsilon_1=0.15625$ (this yields a maximum magnitude of 10 for $E$, compared to 63.97 for $A$), $\epsilon_2=1$ (this was chosen slightly higher than the value corresponding to the maximum spatial derivative for the unperturbed flow), and $\epsilon_3=0.05$ (this means it takes more than six grid boxes for $E$ to switch from its maximum allowed value to its minimum allowed value or vice-versa).
Using $K=6$ and the above parameters, we solve the optimisation problem in Section \ref{lpsection}.
We ran FICO Xpress solver (version 7.9) on a standard desktop/laptop using the barrier method with crossover to achieve a basic feasible solution (a basic solution has the least number of nonzero values in the optimal solution).
The 6 dominant eigenvalues of $A_n$ are $0, -0.0774, -0.1970, -0.3138\pm 1.0484i, -0.3641$.
The 6 dominant eigenvalues of the perturbed generator $A_n+E_n$ are $0, -0.0962, -0.2555, -0.3499\pm 0.9088i, -0.4564\pm 2.2666i$.
Thus, as promised, the spectrum (apart from the spectral value $\lambda=0$ corresponding to the invariant density) has been pushed farther from the imaginary axis.
A comparison between the original and optimized vector fields is shown in Figure \ref{quivers}



We smooth the vector field by applying a cubic smoothing spline to the numerically derived stream function as described in Section \ref{sec:smooth} and then construct the velocity field from the smoothed stream function.
The cubic smoothing spline can be easily implemented using \verb"csaps" in MATLAB, which minimizes a weighted sum of the weighted mean-square error over the data points and a smoothness constraint described by the second derivative of the interpolated function.
We use a uniform data weight on the data points and smoothing parameter $p=0.9925$.

\begin{figure}[H]
  \centering
  \includegraphics[scale=0.45]{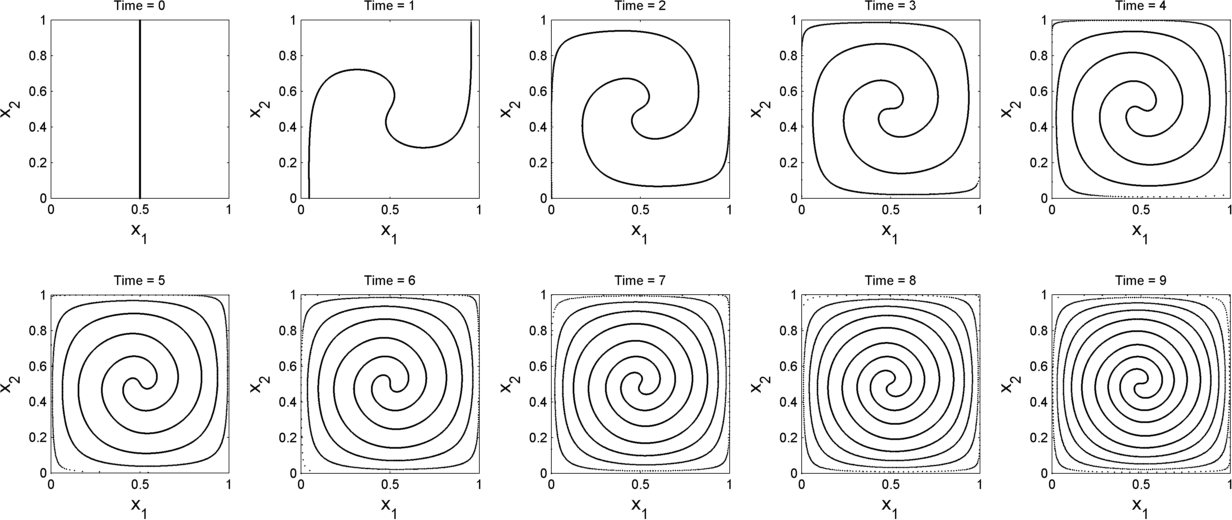}\\
  \caption{The evolution of a vertical line segment under the optimized flow from time $T=0$ to $T=9$.}\label{line_evo2}
\end{figure}


We evolve the curve from Figure \ref{line_evo} by the optimally perturbed vector field; see Figure \ref{line_evo2}.
Consistent with the quantitative improvement of $\Lambda$ from -0.0774 to -0.0962, one sees both a more rapid and more uniform filling of the domain by the evolved curve, when compared with Figure \ref{line_evo}.
The optimisation achieves this in part by increasing slightly the speed near $x=(1/2,1/2)$ and compensating by decreasing speed away from $x=(1/2,1/2)$.
Other compensatory effects ensure that area is preserved by the optimized vector field.

\subsection{Case study 2: Periodic double gyre}
\begin{figure}[H]
  \centering
  \includegraphics[width=\textwidth]{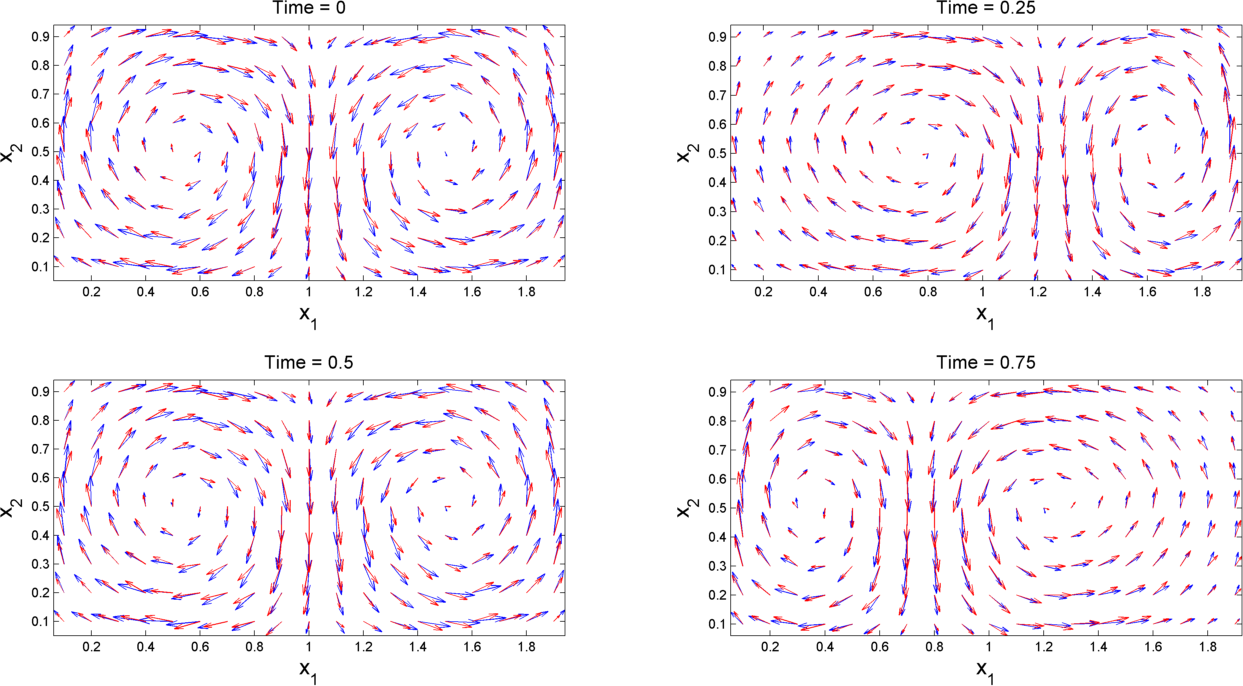}\\
  \caption{Comparison between the original (blue) and optimized (red) velocities of the double-gyre flow at times $t=0, 1/4, 1/2, 3/4$ in the driving cycle of period 1.}\label{quiver_dg}
\end{figure}

\begin{figure}[H]
  \centering
  \includegraphics[width=9cm]{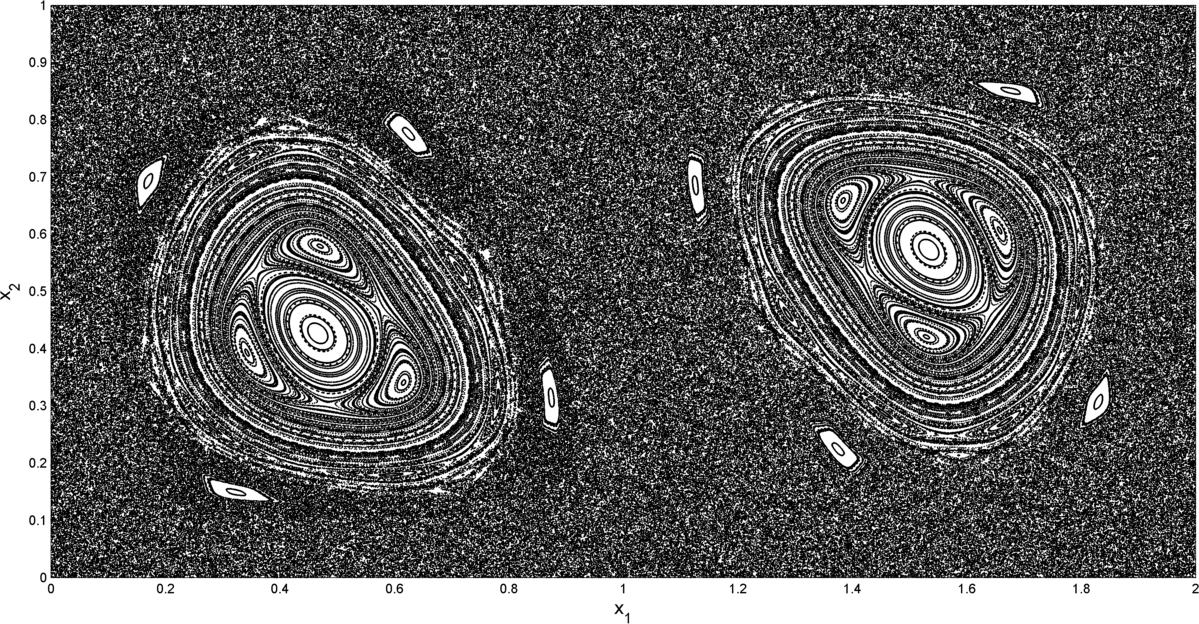}\\
  \caption{The Poincar\'e map at the time slice $t=0$ obtained from 1250 (uniformly distributed) sample points iterated over 500 periods.}\label{poincare}
\end{figure}

\begin{figure}[H]
  \centering
  \includegraphics[width=9cm]{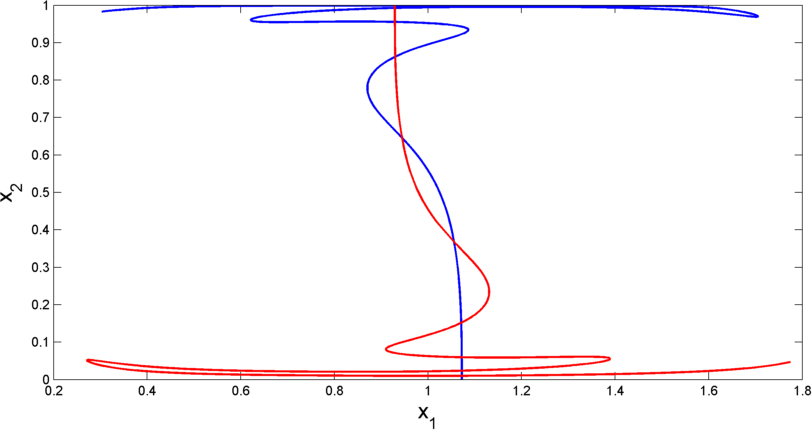}\\
  \caption{Segments of the stable (blue) and unstable (red) manifolds of the double gyre (\ref{dgeqn}).}\label{lobes}
\end{figure}

Our second case study is the periodically forced double gyre on a rectangular domain $M=[0,2]\times [0,1]$, with forcing period of one time unit.
The flow is generated by the time-periodic velocity field
\begin{equation}
\label{dgeqn}
v(t,x)=(-(\pi/4)\sin(\pi f(t,x_1))\cos(\pi x_2),(\pi/4)\cos(\pi f(t,x_1))\sin(\pi x_2)\cdot (df/d{x_1})(t,x_1)),
\end{equation}
where $f(t,x_1)=\sin(2\pi t)x_1^2/4+(1-\sin(2\pi t)/2)x_1$; see Figure \ref{quiver_dg} (blue).
Two gyres are separated by an oscillating vertical separatrix; this oscillation creates complicated fluid flow between the left and right halves of the domain.
It is well known in the dynamical systems and fluid dynamics literature that the double gyre flow appears to exhibit a combination of regular and chaotic behaviour, with KAM tori explaining quasi-periodic behaviour in the regular regions.
The Poincar\'e map at the time slice $t=0$ shown in Figure \ref{poincare} shows this mixed phase space.
An important transport mechanism in the chaotic region is so-called ``lobe dynamics'' \cite{MMP84,romkedar_wiggins_90}, where fluid is transported by ``lobes'' formed by intersections of stable and unstable manifolds of hyperbolic periodic points on the upper and lower horizontal boundaries of $M$.
Parts of these manifolds and some of the major lobes are shown in Figure \ref{lobes}.

\begin{figure}[H]
  \centering
  \includegraphics[width=18cm]{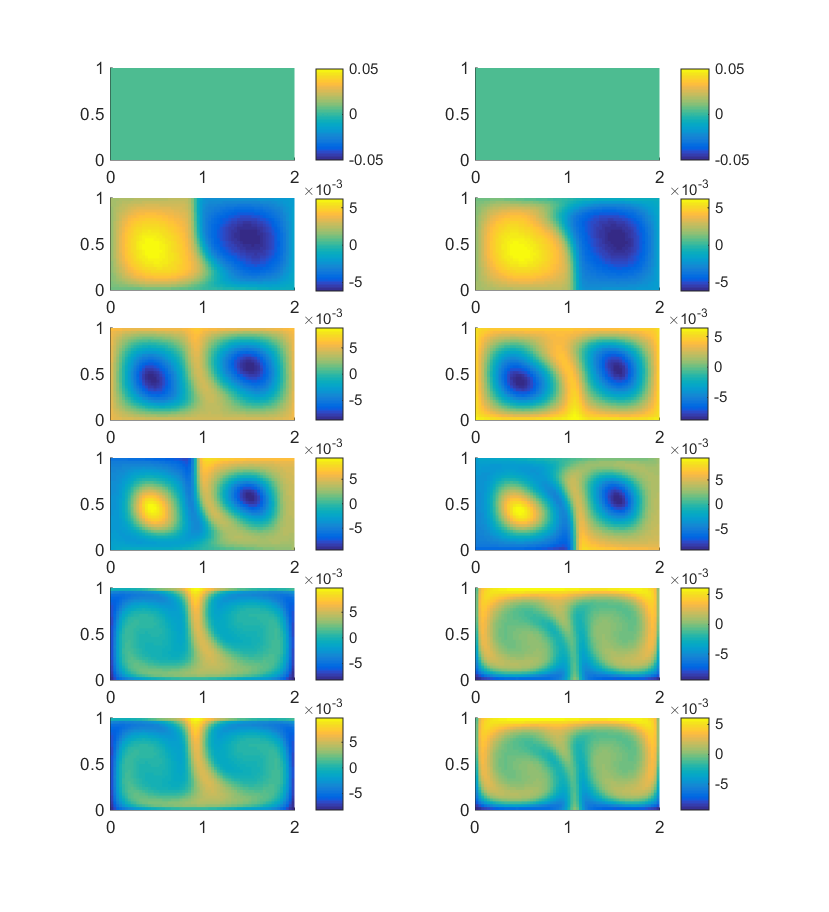}\\
  \caption{Real parts of eigenvectors of $A_n$ corresponding to the eigenvalues $\lambda_1,\ldots,\lambda_6$ in descending order and restricted to the time slice $t=0$.  Upper: $\lambda_1$, lower: $\lambda_6$.  The left column shows left eigenvectors and the right column shows right eigenvectors.}\label{double_gyre_evecs}
\end{figure}

We partition $S^1\times M$ into a grid of $32\times 64\times 32$ identical cubes and use (\ref{ulamformula2}) to create a $2^{16}\times 2^{16}$ matrix $A_n$ approximating the generator $\mathcal{A}$.
The real parts of the leading six left (resp.\ right) eigenvectors are shown in the left (resp.\ right) column of Figure \ref{double_gyre_evecs}.
The corresponding eigenvalues are (in descending order) $0, -0.0483, -0.1746, -0.2947, -0.3148\pm 0.9503i$.
The most important dynamical features are captured by the top three rows of Figure \ref{double_gyre_evecs}.
As in the single gyre case study, the leading eigenvectors are constant, consistent with the fact that the flow is area preserving.
The second row's left column (resp.\ right column) of Figure \ref{double_gyre_evecs} shows a clear separation of the left (yellow) and right (blue) hand sides of $M$, with a thin, green, approximately vertical dividing strip an approximate neighborhood of the unstable manifold (resp.\ stable manifold) of one of the two hyperbolic periodic points. 
We refer the reader to \cite{FPG09} for further details on the hyperbolic periodic points and corresponding unstable/stable manifolds.
The third row of Figure \ref{double_gyre_evecs} shows the separation between the two regular regions (coloured blue) and the surrounding chaotic region (coloured green/yellow).
This indicates the difficulty of transport from the regular regions to the chaotic regions.
For the purely advective flow, the regular regions are composed of invariant sets in extended phase space; thus no transport in or out is possible.
However, with the amount of numerical diffusion in our experiment, the transport in and out of these sets is in fact greater than across the oscillating separatrix.
This explains why the strange eigenmodes corresponding to the regular regions appear below the eigenmodes corresponding to the lobe dynamics when ordered by the real parts of the corresponding eigenvalues.
This ``noise bifurcation'' is observed and discussed further in \cite{FPG14}. The remaining rows of Figure \ref{double_gyre_evecs} show the next most slowly decaying modes.

\begin{figure}[H]
  \centering
  \includegraphics[width=18cm]{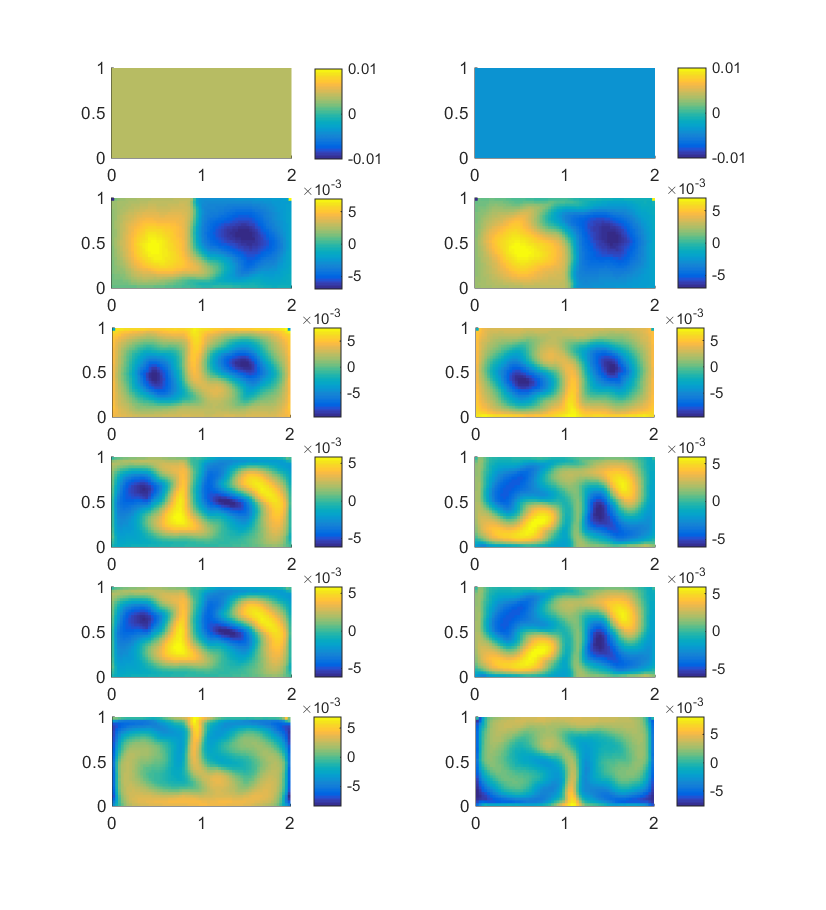}\\
  \caption{Real parts of eigenvectors of $A_n+E$ corresponding to the eigenvalues $\lambda_1,\ldots,\lambda_6$ in descending order and restricted to the time slice $t=0$.  Upper: $\lambda_1$, lower: $\lambda_6$.  The left column shows left eigenvectors and the right column shows right eigenvectors.}\label{double_gyre_evecs_pert}
\end{figure}

We compute $\max_{1\le i\neq j\le n}A_{n,ij}\approx 37.2904$ and $\Delta=1/32=0.03125$.
We choose parameters $\epsilon_1=0.28125$ (this yields a maximum magnitude of 9 for $E$, compared to 37.29 for $A$), $\epsilon_2=1.5$ (this is chosen slightly higher than the value corresponding to the maximum spatial derivative for the unperturbed flow), and $\epsilon_3=0.1$ (this means it takes six grid boxes for $E$ to switch from its maximum allowed value to its minimum allowed value).
Using $K=6$ and the above parameters, we solve the optimisation problem in Section \ref{lpsection}.
The linear programming problem after preprocessing contained 197014 variables (this is approximately $3\times 2^{16})$ and 389574 constraints, and was solved using the primal simplex algorithm on FICO Xpress Optimizer (version 7.9) on a standard desktop/laptop.
The 6 dominant eigenvalues of the perturbed generator $A_n+E_n$ are $0,-0.1007,-0.2787,-0.3814\pm 1.5319i, -0.4078\pm 0.9989i$.
Thus, the spectrum (apart from the spectral value $\lambda=0$ corresponding to the invariant density) has been pushed farther from the imaginary axis.
The corresponding eigenvectors for the perturbed generator are shown in Figure \ref{double_gyre_evecs_pert}.
We note that in this particular experiment the optimizer additionally found eigenvalues $-0.0867$ and $-0.0869$.
The eigenvectors of these eigenvalues at time $t=0$ have supports restricted to the very top left and top right boxes (those containing the points $x=(0,1)$ and $x=(2,1)$).
The likely reason for this is that these ``corner'' boxes have limited freedom in terms of flow through their open two faces;  recall that the net flux must remain zero.
It turns out that in this situation they are the ``mixing bottleneck'' at our current discretisation level.
As these eigenvalues are numerical artifacts we have removed them from the discussion, and the corresponding eigenvectors are not shown in Figure~\ref{double_gyre_evecs_pert}.

Considering the eigenvectors from the perturbed generator in Figure~\ref{double_gyre_evecs_pert}, we see that the leading eigenvectors (top row) are again constant, indicating the area is preserved by the corresponding perturbed flow.
The second and third rows of Figure~\ref{double_gyre_evecs_pert} have similar overall structure to the second and third rows of Figure~\ref{double_gyre_evecs}, however, the shapes of the blue and yellow regions in these strange eigenmodes have been deformed by the perturbation.
In contrast, the fourth row of Figure~\ref{double_gyre_evecs}, corresponding to a real eigenvalue and indicating further KAM orbit detail, has undergone a major structural change, becoming a two-dimensional eigenspace corresponding to a complex-conjugate pair of eigenvalues.
Our interpretation of this phenomenon is that the perturbation has successfully broken up some of the KAM tori.


A comparison between the original and optimized vector fields at the time slice $t=0$ is shown in Figure \ref{quiver_dg}.
One feature of the optimized vector field that may contribute to greater mixing is that it is mostly faster at times $t=0.25, 0.75$ when the separatrix is maximally off-centre, leading to more rapid transport between the left and right hand sides of the domain.
The optimized vector field compensates for this by being slower at  times $t=0, 0.5$ when the separatrix is centred.

\begin{figure}[H]
  \centering
  \includegraphics[width=9cm]{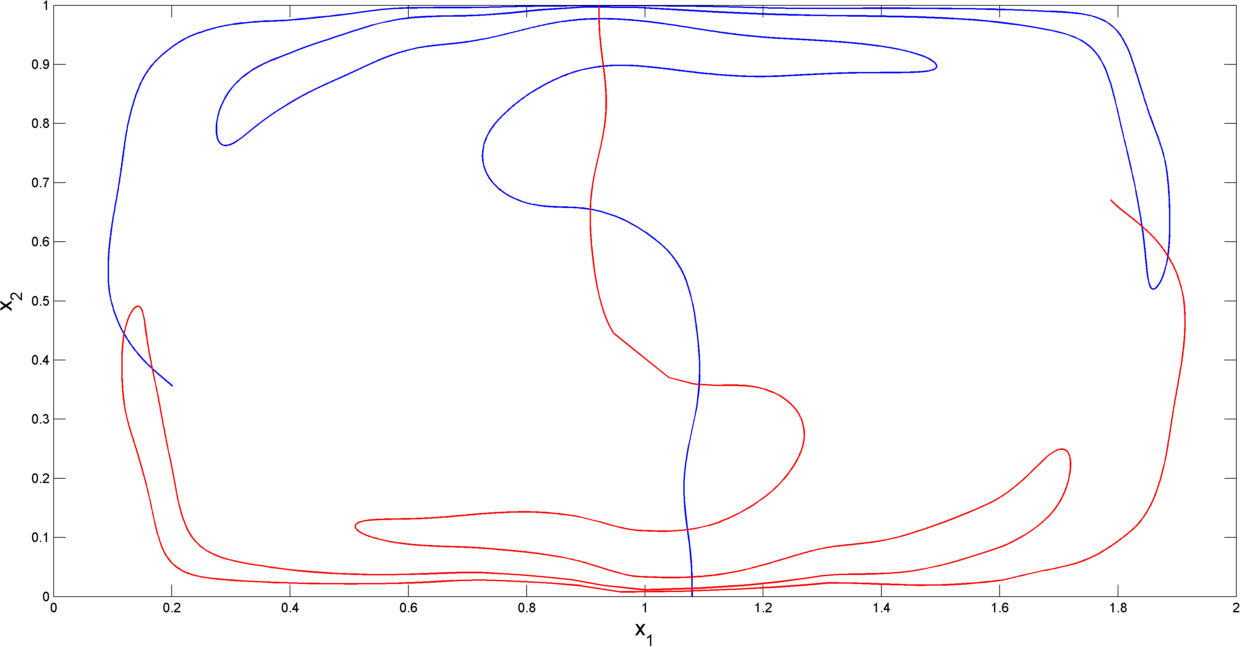}\\
  \caption{Segments of the stable (blue) and unstable (red) manifolds of the perturbed flow.}\label{lobes_perturbed}
\end{figure}

We smooth the vector field as in the previous case study using the same smoothing parameter. 
The flow is not perturbed on the boundary of the domain $M$ and so the hyperbolic periodic points on the boundary remain unchanged.
The stable and unstable manifolds of these periodic points for the optimized flow are shown in Figure \ref{lobes_perturbed}.

\begin{figure}[H]
  \centering
  \includegraphics[width=9cm]{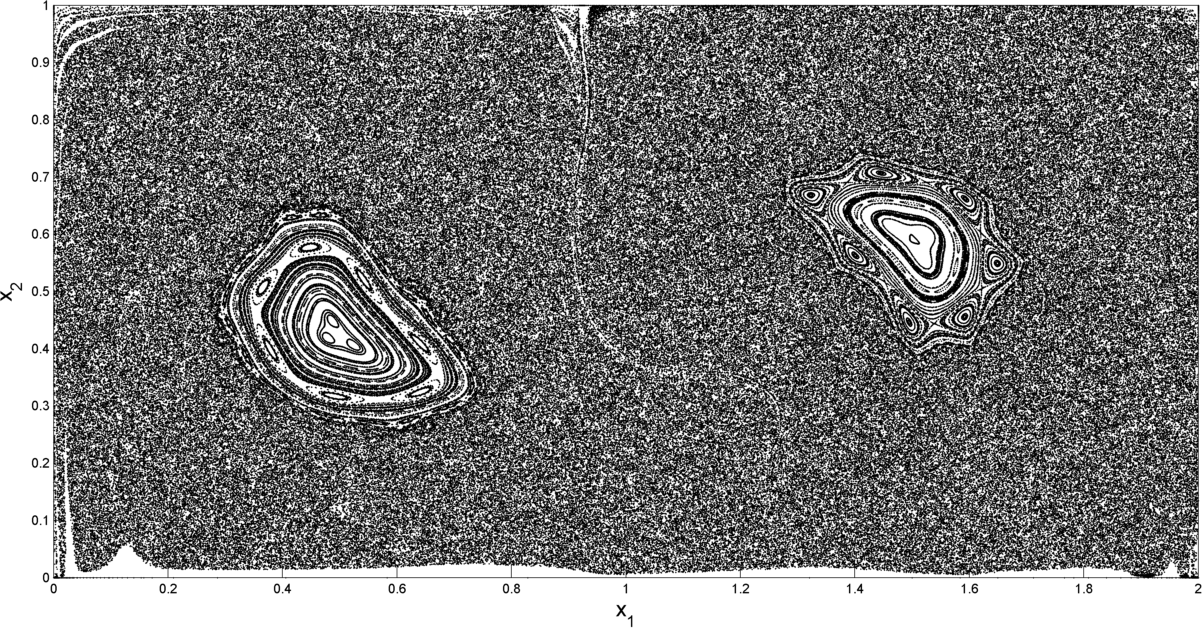}\\
  \caption{The Poincar\'e map at the time slice $t=0$ of the perturbed flow.}\label{poincare_perturbed}
\end{figure}

The lobes for the perturbed flow are clearly much larger than the lobes of the original flow;  thus the transport from the left half of $M$ to the right half of $M$ by lobe dynamics will be significantly enhanced. We emphasise that our spectral optimization approach uses \emph{no specific geometric information}.
That is, the optimization knows nothing of stable or unstable manifolds, nor lobes.
Nevertheless, we can observe that in this experiment the optimizer exploits the existing lobe transport mechanism in order to enhance mixing.
A Poincar\'e map of the perturbed flow is shown in Figure \ref{poincare_perturbed}.
The perturbed flow still has regular regions, however, their extent is significantly reduced relative to the original flow.
We emphasise again that our spectral optimisation method does not have access to geometric information regarding KAM tori, but the resulting perturbed flow manages to break up many of these tori and dramatically reduce the size of the regular region.

\begin{figure}[H]
  \centering
  \includegraphics[width=0.75\textwidth]{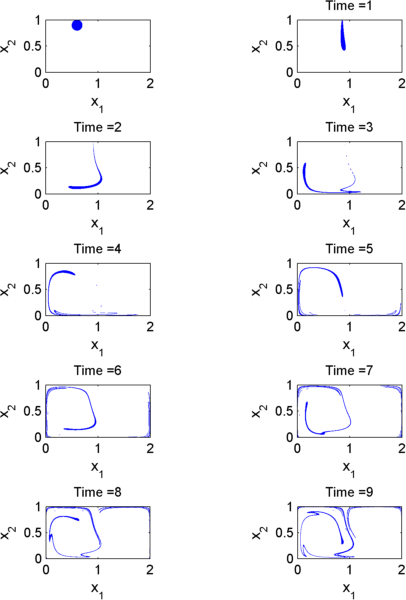}\\
  \caption{Evolution of a disk with the center (0.6,0.9) and radius 1 by the unperturbed flow for 9 forcing periods.}\label{origblob}
\end{figure}

\begin{figure}[H]
  \centering
  \includegraphics[width=0.75\textwidth]{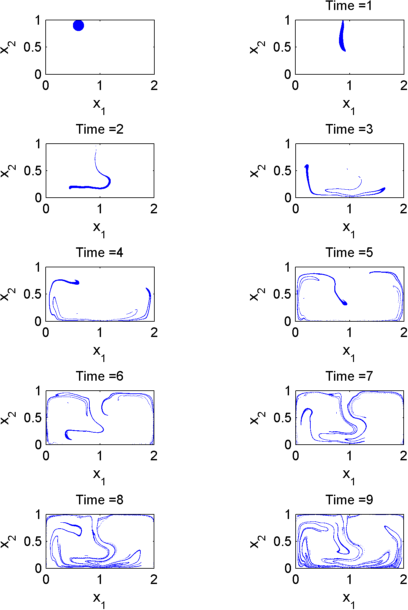}\\
  \caption{Evolution of a disk with the center (0.6,0.9) and radius 1 by the perturbed flow for 9 forcing periods.}\label{pertblob}
\end{figure}

Finally, we demonstrate that the optimized flow mixes a small blob of dye much faster than the original flow;  see Figures \ref{origblob} and \ref{pertblob}.
The perturbed flow evolves the blob so as to distribute the dye over larger portions of the phase space in the same duration of time than the original flow.
 The dye in the optimized flow also penetrates the right half of the domain earlier, and to a much greater extent, than the original flow.
 Both of these observations are consistent with the increase in spectral gap of the generator achieved by our optimization procedure.

\subsection{Optimally inhibiting mixing and controlling specific spatial structures}
One can easily optimally inhibit mixing via local perturbation by replacing the ``min'' with a ``max'' in the objective (\ref{objective}), reversing the inequality in (\ref{multiobj}) and setting $K=2$.
This has the effect of moving the eigenvalue $\lambda_2$ as close as possible to the imaginary axis, subject to the $C^1$ perturbation constraints.

\begin{figure}[H]
  \centering
  \includegraphics[width=8cm]{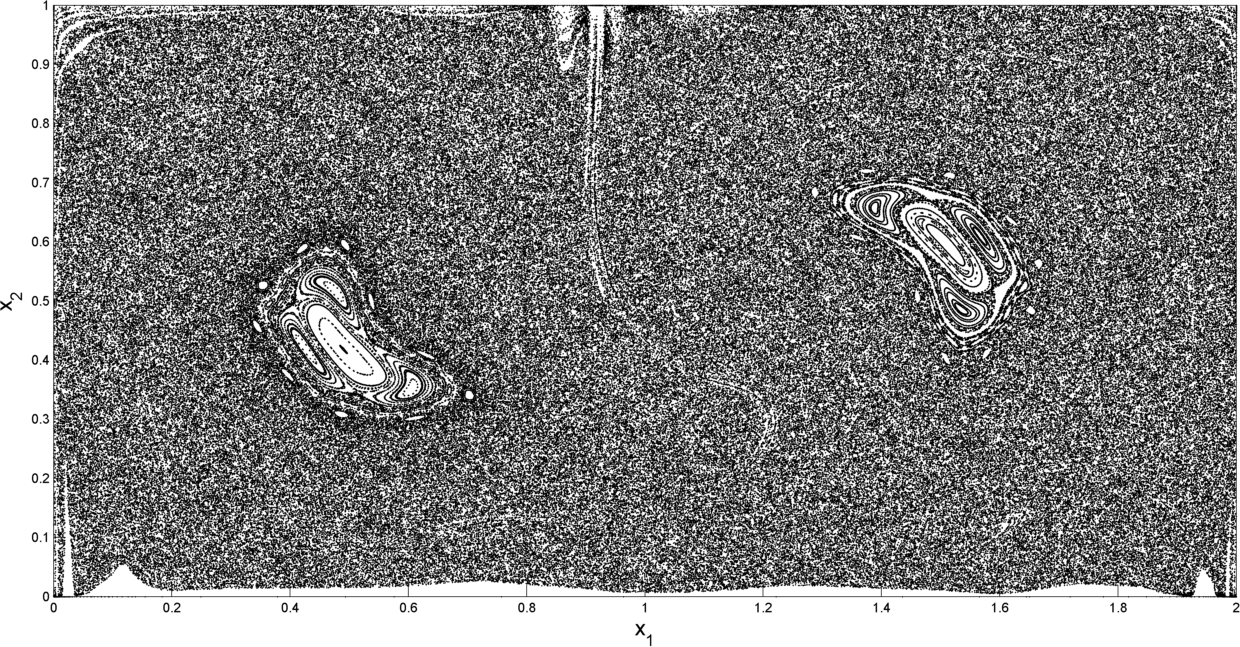}
  \includegraphics[width=8cm]{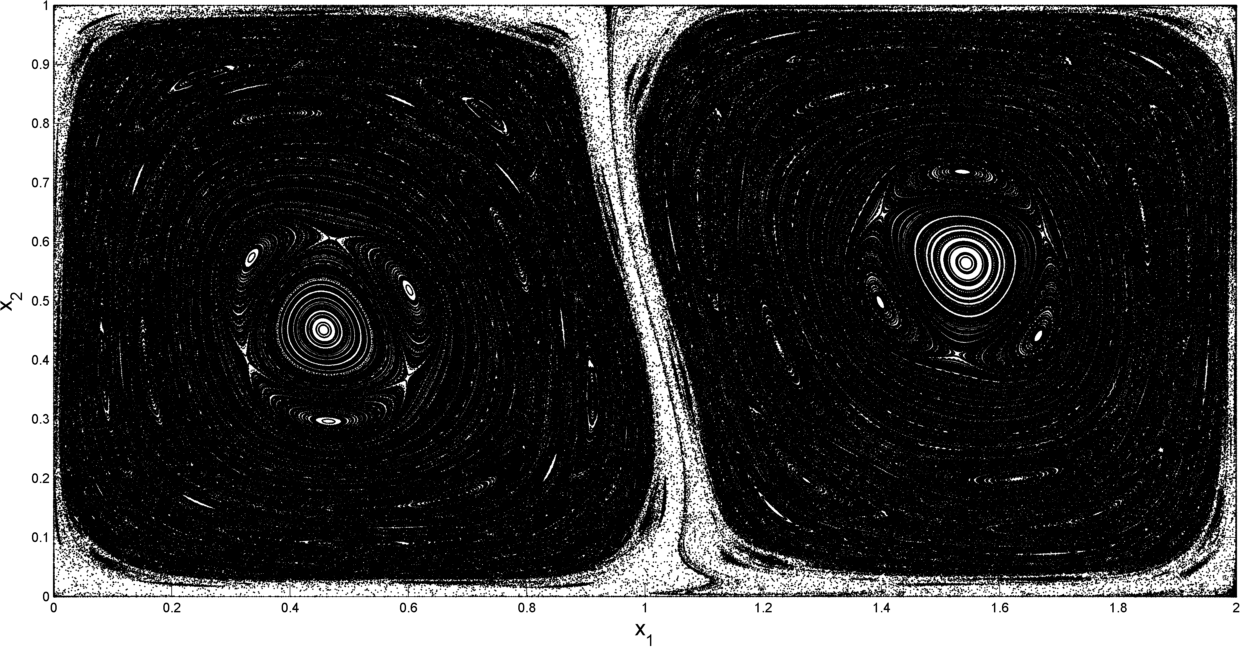}
  \caption{Poincar\'e map at the time slice $t=0$ obtained 1250 (uniformly distributed) sample points iterated over 500 periods. Left: destroying the regular regions as much as possible;  Right: Enhancing the regular regions as much as possible.}\label{poincare3}
\end{figure}

Furthermore, one can also target specific structures highlighted by the strange eigenmodes given by the leading eigenfunctions.
For example, suppose that instead of enhancing global mixing, one instead wishes to destroy  the regular regions of the double gyre as much as possible.
These regular regions are highlighted by the eigenfunction corresponding to $\lambda_3$ and so to achieve this aim, one would solve the linear program (\ref{objective})--(\ref{casemaxderivpert2}), changing (\ref{multiobj}) to include only the term $k=3$.
This linear program pushes $\lambda_3$ away from the imaginary axis, increasing the mixing rate associated with the regular regions.
The result, using the same settings as discussed previously, is shown in Figure \ref{poincare3} (left).
Note that the regular regions are now slightly smaller than in Figure \ref{poincare_perturbed}, consistent with our new objective.

Alternatively, one may wish to enhance the regular regions as much as possible by local perturbation.
One can replace the ``min'' with a ``max'' in the the objective (\ref{objective}), reverse the inequality in (\ref{multiobj}) and include only the term $k=3$ in (\ref{multiobj}).
The resulting linear program pushes $\lambda_3$ as close to the imaginary axis as is allowed by the $C^1$ neighborhood.
The outcome of this optimization is shown in Figure \ref{poincare_perturbed}, where the regular regions have dramatically increased in size to almost fill the entire domain.

\section{Conclusion}

We introduced a numerical method to optimally enhance the mixing rate of a flow through perturbation of the underlying vector field.
Specifically, given a $C^1$ neighborhood of the current vector field, we found the vector field in this neighborhood that has the most rapid mixing rate.
Our approach was to construct a discrete generator matrix for the vector field, thus achieving a linear representation of the nonlinear flow.
An approximate mixing rate for the flow is now also accessible via the second eigenvalue of the generator matrix.
To increase the rate of mixing, we use a linear program to perturb the generator matrix to push the nontrivial spectrum of the generator as far as possible from the imaginary axis.
From the solution of the linear program, we then inferred the optimized vector field.

To our knowledge, this is the first practical way to perturb the entire vector field to increase the rate of mixing.
Previous approaches have tuned a small number of parameters, or switched between a handful of prespecified vector fields.
Our technique shows what mixing rates are achievable if a full optimization of the vector field is performed.

Our approach can be applied to both steady and periodically driven flows, and is very flexible from the point of view of enforcing practical constraints.
For example, certain regions of the phase space or the driving period can easily be excluded from the perturbation (i.e.\ no perturbation allowed).
Moreover, the technique can be trivally adapted to instead slow down the rate of mixing, or to destroy or enhance particular features of the flow.

Straightforward extensions of our approach include the simultaneous optimization of the diffusion term in advection-diffusion equations, and the incorporation of our method as a single step in a descent algorithm if larger perturbations are allowed.

%
%

\section{Acknowledgements}
GF is supported by an ARC Future Fellowship.
NS is supported by the Department of Mathematics, University of Surrey.
Both authors thank two anonymous referees for helpful comments.

\bibliographystyle{abbrv}

\begin{thebibliography}{10}

\bibitem{ACM}
G.~Alberti, G.~Crippa, and A.~L. Mazzucato.
\newblock Exponential self-similar mixing by incompressible flows.
\newblock {\em arXiv preprint arXiv:1605.02090}, 2016.

\bibitem{ChaoticAdvectionReview}
H.~Aref et~al.
\newblock {Frontiers of chaotic advection}.
\newblock arXiv:1403.2953, 2014.

\bibitem{Balasuriya15}
S.~Balasuriya.
\newblock Dynamical systems techniques for enhancing microfluidic mixing.
\newblock {\em Journal of Micromechanics and Microengineering}, 25:094005,
  2015.

\bibitem{Batchelorbook}
G.~K. Batchelor.
\newblock {\em An introduction to fluid dynamics}.
\newblock Cambridge University Press, 1967.

\bibitem{Cortelezzi_Adrover_Giona08}
L.~Cortelezzi, A.~Adrover, and M.~Giona.
\newblock Feasibility, efficiency and transportability of short-horizon optimal
  mixing protocols.
\newblock {\em J. Fluid Mech.}, 597:199--231, 2008.

\bibitem{DFS00}
M.~Dellnitz, G.~Froyland, and S.~Sertl.
\newblock On the isolated spectrum of the {P}erron-{F}robenius operator.
\newblock {\em Nonlinearity}, 13(4):1171--1188, 2000.

\bibitem{DellnitzJunge99}
M.~Dellnitz and O.~Junge.
\newblock On the approximation of complicated dynamical behavior.
\newblock {\em SIAM J. Numer. Anal.}, 36(2):491--515, 1999.

\bibitem{FH04}
D.~R. Fereday and P.~H. Haynes.
\newblock Scalar decay in two-dimensional chaotic advection and
  {B}atchelor-regime turbulence.
\newblock {\em Phys. of Fluids}, 16(12):4359--4370, 2004.

\bibitem{F07}
G.~Froyland.
\newblock On {U}lam approximation of the isolated spectrum and eigenfunctions
  of hyperbolic maps.
\newblock {\em Disc. and Cont. Dyn. Sys., Series A}, 17(3):671--689, 2007.

\bibitem{FGTW}
G.~Froyland, C.~Gonz\'{a}lez-Tokman, and T.~Watson.
\newblock Optimal mixing enhancement by local perturbation.
\newblock {\em To appear in SIAM Review}, 2016.

\bibitem{FroylandAgulhas}
G.~Froyland, C.~Horenkamp, V.~Rossi, N.~Santitissadeekorn, and A.~S. Gupta.
\newblock {Three-dimensional characterization and tracking of an Agulhas Ring}.
\newblock {\em Ocean Modelling}, 52--53(0):69 -- 75, 2012.

\bibitem{FJK13}
G.~Froyland, O.~Junge, and P.~Koltai.
\newblock Estimating long-term behavior of flows without trajectory
  integration: The infinitesimal generator approach.
\newblock {\em SIAM Journal on Numerical Analysis}, 51(1):223--247, 2013.

\bibitem{FrKo}
G.~Froyland and P.~Koltai.
\newblock Estimating long-term behavior of periodically driven flows without
  trajectory integration.
\newblock arXiv, http://arxiv.org/abs/1511.07272, 2015.

\bibitem{FroylandLloydSanti}
G.~Froyland, S.~Lloyd, and N.~Santitissadeekorn.
\newblock Coherent sets for nonautonomous dynamical systems.
\newblock {\em Physica D}, 239(16):1527--1541, 2010.

\bibitem{FPET}
G.~Froyland, K.~Padberg, M.~England, and A.~Treguier.
\newblock Detection of coherent oceanic structures via transfer operators.
\newblock {\em Physical review letters}, 98(22):224503, 2007.

\bibitem{FPG09}
G.~Froyland and K.~Padberg-Gehle.
\newblock Almost-invariant sets and invariant manifolds-connecting
  probabilistic and geometic descriptions of coherent structures in flows.
\newblock {\em Physica D}, 23(16):1507--1523, 2009.

\bibitem{FPG14}
G.~Froyland and K.~Padberg-Gehle.
\newblock Almost-invariant and finite-time coherent sets: Directionality,
  duration, and diffusion.
\newblock In W.~Bahsoun, C.~Bose, and G.~Froyland, editors, {\em Ergodic
  Theory, Open Dynamics, and Coherent Structures}, pages 171--216. Springer,
  New York, NY, 2014.

\bibitem{FroylandSantiMonahan}
G.~Froyland, N.~Santitissadeekorn, and A.~Monahan.
\newblock Transport in time-dependent dynamical systems: Finite-time coherent
  sets.
\newblock {\em Chaos}, 20(4):043116, 2010.

\bibitem{FSvS14}
G.~Froyland, R.~M. Stuart, and E.~van Sebille.
\newblock How well-connected is the surface of the global ocean?
\newblock {\em Chaos}, 24(3):033126, 2014.

\bibitem{GorodetskyiSpeetjensAnderson}
O.~Gorodetskyi, M.~F.~M. Speetjens, and P.~D. Anderson.
\newblock An efficient approach for eigenmode analysis of transient
  distributive mixing by the mapping method.
\newblock {\em Physics of Fluids}, 24(5):053602, 2012.

\bibitem{GouillartDauchotThiffeault}
E.~Gouillart, O.~Dauchot, and J.-L. Thiffeault.
\newblock Measures of mixing quality in open flows with chaotic advection.
\newblock {\em Physics of Fluids}, 23(1):013604, 2011.

\bibitem{GouillartDauchotThiffeaultRoux}
E.~Gouillart, O.~Dauchot, J.-L. Thiffeault, and S.~Roux.
\newblock Open-flow mixing: Experimental evidence for strange eigenmodes.
\newblock {\em Physics of Fluids}, 21(2):023603, 2009.

\bibitem{Gubanov_Cortezzi10}
O.~Gubanov and L.~Cortelezzi.
\newblock On the cost efficiency of mixing optimization.
\newblock {\em Journal of Fluid Mechanics}, 692:112--136, 2012.

\bibitem{HV05}
P.~H. Haynes and J.~Vanneste.
\newblock What controls the decay of passive scalars in smooth flows?
\newblock {\em Phys. Fluids}, 17(097103), 2005.

\bibitem{hornjohnson}
R.~A. Horn and C.~R. Johnson.
\newblock {\em Matrix analysis}.
\newblock Cambridge University Press, 2nd edition, 2012.

\bibitem{IKX}
G.~Iyer, A.~Kiselev, and X.~Xu.
\newblock Lower bounds on the mix norm of passive scalars advected by
  incompressible enstrophy-constrained flows.
\newblock {\em Nonlinearity}, 27(5):973, 2014.

\bibitem{katok}
A.~Katok and B.~Hasselblatt.
\newblock {\em Introduction to the modern theory of dynamical systems}.
\newblock Cambridge University Press, 1997.

\bibitem{LasotaMackey}
A.~Lasota and M.~C. Mackey.
\newblock {\em Chaos, fractals, and noise: stochastic aspects of dynamics},
  volume~97.
\newblock Springer, 2013.

\bibitem{leveque}
R.~J. LeVeque.
\newblock {\em Finite volume methods for hyperbolic problems}, volume~31.
\newblock Cambridge University Press, 2002.

\bibitem{LinThiffeaultDoering}
Z.~Lin, J.-L. Thiffeault, and C.~R. Doering.
\newblock Optimal stirring strategies for passive scalar mixing.
\newblock {\em Journal of Fluid Mechanics}, 675:465--476, 2011.

\bibitem{LiuHaller}
W.~Liu and G.~Haller.
\newblock Strange eigenmodes and decay of variance in the mixing of diffusive
  tracers.
\newblock {\em Physica D: Nonlinear Phenomena}, 188(1--2):1 -- 39, 2004.

\bibitem{LLetal}
E.~Lunasin, Z.~Lin, A.~Novikov, A.~Mazzucato, and C.~R. Doering.
\newblock Optimal mixing and optimal stirring for fixed energy, fixed power, or
  fixed palenstrophy flows.
\newblock {\em Journal of Mathematical Physics}, 53(11):115611, 2012.

\bibitem{MMP84}
R.~S. Mackay, J.~D. Meiss, and I.~C. Percival.
\newblock Transport in {H}amiltonian systems.
\newblock {\em Physica D}, 13:55--81, 1984.

\bibitem{Mathew_Mezic_Grivopoulos_Vaidya_Petzold}
G.~Mathew, I.~Mezic, S.~Grivopoulos, U.~Vaidya, and L.~Petzol{d}.
\newblock Optimal control of mixing in stokes fluid flows.
\newblock {\em J. Fluid Mech.}, 580:261--281, 2007.

\bibitem{MathewMezicPetzold}
G.~Mathew, I.~Mezi{\'c}, and L.~Petzold.
\newblock A multiscale measure for mixing.
\newblock {\em Phys. D}, 211(1-2):23--46, 2005.

\bibitem{Padberg-Gehle_Ober-Blobaum}
S.~Ober-Bl\"{o}baum and K.~Padberg-Gehle.
\newblock Multiobjective optimal control of fluid mixing.
\newblock {\em PAMM}, 15(1):639--640, 2015.

\bibitem{Pazy}
A.~Pazy.
\newblock {\em Semigroups of Linear Operators and Applications to Partial
  Differential Equations}, volume~44 of {\em Applied Mathematical Sciences}.
\newblock Springer-Verlag New York, 1993.

\bibitem{Petersen}
K.~Petersen.
\newblock {\em Ergodic Theory}, volume~2 of {\em Cambridge Studies in Advanced
  Mathematics}.
\newblock Cambridge University Press, 1990.

\bibitem{Pierrehumbert94}
R.~Pierrehumbert.
\newblock Tracer microstructure in the large-eddy dominated regime.
\newblock {\em Chaos, Solitons \& Fractals}, 4(6):1091 -- 1110, 1994.
\newblock Special Issue: Chaos Applied to Fluid Mixing.

\bibitem{PikovskyPopovych}
A.~Pikovsky and O.~Popovych.
\newblock Persistent patterns in deterministic mixing flows.
\newblock {\em EPL (Europhysics Letters)}, 61(5):625, 2003.

\bibitem{Provenzale}
A.~Provenzale.
\newblock Transport by coherent barotropic vortices.
\newblock {\em Annual Review of Fluid Mechanics}, 31(1):55--93, 1999.

\bibitem{romkedar_wiggins_90}
V.~Rom-Kedar and S.~Wiggins.
\newblock Transport in two-dimensional maps.
\newblock {\em Archive for Rational Mechanics and Analysis}, 109:239--298,
  1990.

\bibitem{SinghKangMeijerAnderson}
M.~K. Singh, T.~G. Kang, H.~E.~H. Meijer, and P.~D. Anderson.
\newblock The mapping method as a toolbox to analyze, design, and optimize
  micromixers.
\newblock {\em Microfluidics and Nanofluidics}, 5(3):313--325, 2008.

\bibitem{SinghSpeetjensAnderson}
M.~K. Singh, M.~F.~M. Speetjens, and P.~D. Anderson.
\newblock Eigenmode analysis of scalar transport in distributive mixing.
\newblock {\em Physics of Fluids}, 21(9):093601, 2009.

\bibitem{SpencerWiley}
R.~S. Spencer and R.~M. Wiley.
\newblock The mixing of very viscous liquids.
\newblock {\em Journal of Colloid Science}, 6(2):133 -- 145, 1951.

\bibitem{SturmanEtAl}
R.~Sturman, J.~M. Ottino, and S.~Wiggins.
\newblock {\em The mathematical foundations of mixing}, volume~22 of {\em
  Cambridge Monographs on Applied and Computational Mathematics}.
\newblock Cambridge University Press, Cambridge, 2006.

\bibitem{ThiffeaultChapter}
J.-L. Thiffeault.
\newblock Scalar decay in chaotic mixing.
\newblock In J.~Weiss and A.~Provenzale, editors, {\em Transport and Mixing in
  Geophysical Flows}, volume 744 of {\em Lecture Notes in Physics}, pages
  3--36. Springer Berlin Heidelberg, 2008.

\bibitem{Thiffeault}
J.-L. Thiffeault.
\newblock Using multiscale norms to quantify mixing and transport.
\newblock {\em Nonlinearity}, 25(2):R1, 2012.

\bibitem{ThiffeaultDoeringGibbon}
J.-L. Thiffeault, C.~R. Doering, and J.~D. Gibbon.
\newblock A bound on mixing efficiency for the advection-diffusion equation.
\newblock {\em Fluid Mechanics}, 521:105--114, 2004.

\bibitem{ThiffeaultPavliotis}
J.-L. Thiffeault and G.~A. Pavliotis.
\newblock Optimizing the source distribution in fluid mixing.
\newblock {\em Physica D: Nonlinear Phenomena}, 237(7):918 -- 929, 2008.

\bibitem{Ulam}
S.~M. Ulam.
\newblock {\em A collection of mathematical problems}.
\newblock Interscience Tracts in Pure and Applied Mathematics, no. 8.
  Interscience Publishers, New York-London, 1960.

\bibitem{Voth}
G.~A. Voth, G.~Haller, and J.~P. Gollub.
\newblock Experimental measurements of stretching fields in fluid mixing.
\newblock {\em Physical review letters}, 88(25):254501, 2002.

\bibitem{Walters}
P.~Walters.
\newblock {\em An Introduction to Ergodic Theory}, volume~79 of {\em Graduate
  Texts in Mathematics}.
\newblock Springer-Verlag New York, 1982.

\bibitem{YZ}
Y.~Yao and A.~Zlatos.
\newblock Mixing and un-mixing by incompressible flows.
\newblock {\em arXiv preprint arXiv:1407.4163}, 2014.

\end{thebibliography}

\end{document}